\pdfoutput=1
\documentclass[reprint,amsmath,amssymb,aps,prb,longbibliography,floatfix]{revtex4-2}
\usepackage{dcolumn}
\usepackage{comment}
\usepackage{amsmath}
\allowdisplaybreaks
\usepackage{bm}
\usepackage{mathrsfs}
\usepackage{here}
\usepackage{mathtools}
\usepackage{physics}
\usepackage{color}
\usepackage{xcolor}
\usepackage{wrapfig}
\usepackage{booktabs}
\usepackage{url}
\usepackage[pdftex]{graphicx}
\usepackage[english]{babel}

\begin{document}

\preprint{APS/123-QED}

\title{Nonequilibrium Quasiparticle Effects \\ on
Domain Wall Dynamics in Superconductors}

\author{Takuma Kanakubo}
\email{kanakubo@vortex.c.u-tokyo.ac.jp}
\affiliation{Department of Physics, The University of Tokyo, Bunkyo-ku, Tokyo 113-0033, Japan}

\author{Taira Kawamura}
\affiliation{Department of Physics, College of Science and Technology, Nihon University, Chiyoda, Tokyo 101-8308, Japan}

\author{Yusuke Kato}
\email{yusuke@phys.c.u-tokyo.ac.jp}
\affiliation{Department of Physics, The University of Tokyo, Bunkyo-ku, Tokyo 113-0033, Japan}
\affiliation{Department of Basic Science, The University of Tokyo, Meguro, Tokyo 153-8902, Japan}

\begin{abstract}
We study the dynamics of a domain wall (DW) in a type-II superconductor
connected to two heat reservoirs.
We employ a generalized time-dependent Ginzburg--Landau framework in which the superconducting
order parameter and the nonequilibrium quasiparticle distribution are treated as coupled dynamical
variables. Within this
framework, the effect of the thermal bias is imposed through boundary conditions on the
quasiparticle distribution, which are set by the reservoir temperatures. We show,
both numerically and within linear response, that the DW moves toward the hotter boundary. From the local momentum-balance relation implied by
the model, we identify a viscous force and a force arising from the
coupling between the order parameter and the nonequilibrium
distribution function. The nonequilibrium distribution separates exactly into a boundary-driven part and a part generated by the motion of the DW itself. The former sets the sign of the DW velocity, whereas the latter renormalizes the relaxation of the order parameter. These results provide
a microscopic basis for the phenomenological local-temperature description developed in our
previous work.
\end{abstract}
\maketitle

\section{Introduction} \label{sec:intro}

Quantized vortices are well-known topological defects in type-II
superconductors. Domain walls (DWs), across which the superconducting order
parameter changes sign, are another class of topological defects in
superconducting systems~\cite{richard2016heat}. A DW of the type considered here arises from the periodic
modulation of the order parameter in the Fulde--Ferrell--Larkin--Ovchinnikov
(FFLO) state~\cite{FF,LO}, for which experimental evidence has been reported in type-II superconductors~\cite{bianchi2003possible,mayaffre2014evidence,kasahara2020evidence}. A temperature
gradient or spin accumulation can drive these defects~\cite{OtterSolomon66, Stephen66, adachi2024timedependent, Kanakubo2026}. Understanding their
nonequilibrium dynamics is important both for elucidating fundamental aspects of
nonequilibrium superconductivity and for developing methods to manipulate them
in superconducting devices~\cite{Veshchunov2016}.

In our recent study~\cite{Kanakubo2026}, we investigated the response of
topological defects in superconductors to heat currents by solving a
time-dependent Ginzburg--Landau (TDGL) equation coupled to a thermal-diffusion
equation. We found that both an isolated DW and an isolated vortex move toward
regions of higher temperature, where the order parameter is locally suppressed.
This direction is opposite to Stephen's classical thermal force on a vortex
line, which points toward the colder region~\cite{Stephen66}. This apparent
discrepancy, however, does not imply a contradiction, because Stephen's argument
applies to a dense array of interacting vortices, whereas
Ref.~\cite{Kanakubo2026} and the present work consider a single isolated
defect.

In Ref.~\cite{Kanakubo2026}, the thermal effect was introduced through a local
temperature field $T(x,t)$, obtained by solving a thermal-diffusion equation
with a phenomenological interpolation for the thermal conductivity. Such a
local-temperature description is justified when quasiparticle energy relaxation
is fast enough to maintain local thermal
equilibrium~\cite{Gorkov&Eliashberg1968}. Near a DW or a vortex core, however,
the order parameter varies over the coherence length, and quasiparticles may not
relax rapidly enough to maintain this local equilibrium. The resulting
energy-resolved correction to the order-parameter equation can thus become
significant~\cite{eliashberg1969nonstationary}.

A natural way to describe this correction is to go beyond a local-temperature
description and retain the nonequilibrium quasiparticle distribution function
explicitly. The resulting formulations are often referred to as generalized or
anomalous TDGL equations and are reviewed in standard references on
nonequilibrium superconductivity~\cite{cyrot1973ginzburg,rammer2011noneq}. An
early formulation was given by Eliashberg for a superconductor with a low
concentration of paramagnetic impurities~\cite{eliashberg1969nonstationary}.
Similar equations have since been applied to vortex viscosity and flux
flow~\cite{GK1971viscous,Hu&Thompson1973,GK1973}, the transport entropy of a
vortex in superconductors with paramagnetic
impurities~\cite{deLange1974,Hu1976,kopnin1975heat}, and phase-slip dynamics and
other nonequilibrium problems in dirty superconductors near
$T_{\textrm{c}}$~\cite{Zhu1998,vodolazov2007enhancement,Vodolazov2010}.

In this work, we use such a generalized TDGL framework to reexamine the DW
motion studied in Ref.~\cite{Kanakubo2026}, where the thermal effect was modeled
by a local temperature field and phenomenological transport coefficients. In
particular, we employ the formulation introduced by Kramer and Watts-Tobin for
current-driven phase-slip
centers~\cite{KramerWattsTobin1978,watts1981nonequilibrium}, hereafter referred
to as the KWT framework. This framework treats the nonequilibrium quasiparticle
distribution function as an independent dynamical variable coupled
self-consistently to the order parameter and to the spectral functions
determined by the Usadel equation~\cite{usadel1970generalized}. The treatment of
quasiparticle relaxation in the KWT framework is based on Eliashberg's inelastic
electron--phonon collision integral for nonequilibrium
superconductors~\cite{eliashberg1972inelastic}. We apply this framework to a DW
driven purely by a temperature difference between the two boundaries of the
sample.

In the standard parametrization, the nonequilibrium quasiparticle distribution
function is decomposed into energy and charge
modes~\cite{Schmid66,schmidschon1975linearized,tinkham1996introduction,
E.W.Ora1978,entin1978mode,entin1979Boltzmann,belzig1999quasiclassical,
Golubov_and_Koshelev2011}. In the present setup, however, there is no transport
current, voltage bias, or phase gradient of the order parameter, and the charge
mode decouples. The terms that would generate the charge mode are proportional to
the superfluid momentum and to the time derivative of the phase, and both vanish
for a real order parameter. We thus focus on the energy-mode correction and show
that it decomposes into a boundary-driven component, which drives the DW, and a
motion-induced component, which renormalizes the relaxation of the order
parameter.

We restrict the analysis to a DW and leave the vortex problem for future work.
This choice avoids several complications specific to vortices. A circulating
supercurrent requires a self-consistent treatment of Amp\`ere's law. The nonzero
phase gradient couples the energy and charge modes. The order parameter also
varies in two transverse directions, which increases the dimensionality of the
problem. Nevertheless, Ref.~\cite{Kanakubo2026} showed within a phenomenological
TDGL framework that an isolated vortex moves in the same direction as an
isolated DW and obeys a comparable velocity formula. The present microscopic
treatment of the DW should therefore provide useful insight into vortex dynamics
as well. We note that a similar strategy was used in Ref.~\cite{stone1996} for
Andreev bound states, where a DW configuration was analyzed before the
corresponding vortex problem.

For this DW problem, both numerical simulations and a linear-response
calculation show that the DW moves toward the hotter boundary even when
nonequilibrium quasiparticle dynamics is treated explicitly. This direction
agrees with that found in Ref.~\cite{Kanakubo2026}, providing microscopic
support for the earlier phenomenological result.

The remainder of this paper is organized as follows. Section~\ref{sec:model}
introduces the model. Section~\ref{sec_num} presents numerical
solutions. Section~\ref{sec:linear} derives the DW velocity from a
linear-response calculation and determines its sign.
Section~\ref{sec:discussion} discusses the results and their relation
to Ref.~\cite{Kanakubo2026}. Section~\ref{sec:summary} gives the conclusion.

\section{Model} \label{sec:model}

\subsection{System}

\begin{figure}[t!]
\includegraphics[width=\linewidth]{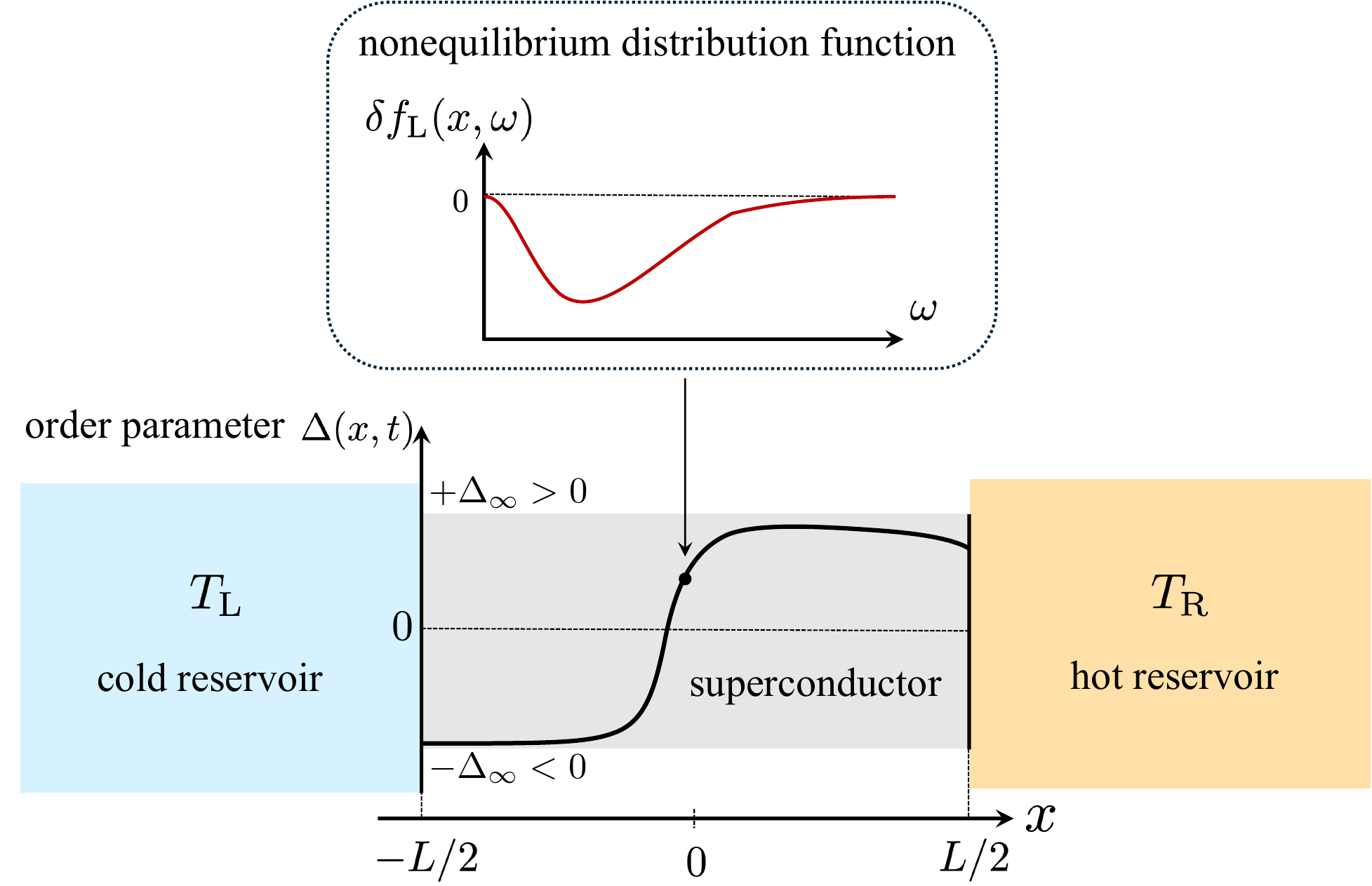}
\caption{Schematic picture of the setup. The superconductor of
length $L$ is connected to a cold reservoir at temperature
$T_{\textrm{L}}$ (left) and a hot reservoir at temperature
$T_{\textrm{R}}$ (right). The order parameter $\Delta(x,t)$ (vertical
axis) changes sign between the two boundaries, forming a DW. The
inset shows, schematically, the nonequilibrium part of the
distribution function $\delta f_{\textrm{L}}(x,\omega)$ as a function
of the quasiparticle energy $\omega$ at the position indicated by the
arrow.}
\label{figsystem}
\end{figure}

We study the nonequilibrium dynamics of a DW in a superconductor
using the generalized TDGL framework described in Sec.~\ref{sec:intro}.
Figure~\ref{figsystem} shows a schematic of the system.

Although the superconductor is three-dimensional, we assume that all physical
quantities vary only along the $x$-direction and are uniform in the $y$- and
$z$-directions. The system has a finite length $L$ along the $x$ direction, with $L$ much larger than the coherence length. At the boundaries
$x = \pm L/2$, the system is in contact with large heat reservoirs held at
temperatures $T_{\textrm{L}}$ (at $x = -L/2$) and $T_{\textrm{R}}$ (at
$x = +L/2$). We assume $T_{\textrm{L}} < T_{\textrm{R}} < T_{\textrm{c}}$, where $T_{\textrm{c}}$ is the critical temperature.

The difference between the two reservoir temperatures, together with
inelastic electron--phonon scattering in the bulk, produces a nonequilibrium quasiparticle population inside the superconductor. In general, such a distribution has two modes, the energy mode and the charge mode~\cite{Schmid66,tinkham1996introduction}. Here we focus on the energy mode alone.

We now clarify how temperature is treated in our formulation. Although the system is
in a nonequilibrium state, the material parameters of the superconductor are
characterized by a single temperature $T$. We describe the nonequilibrium state
solely through the nonequilibrium distribution function, and thus we do not introduce or solve an equation for a local temperature $T(x)$. The reservoir temperatures, $T_{\textrm{L}}$ and $T_{\textrm{R}}$, enter the model only through the boundary conditions, where
they specify the quasiparticle energy distribution at each boundary. We consider a sufficiently small temperature difference, $[T_{\textrm{R}} - T_{\textrm{L}}]/T_{\textrm{c}} \ll 1$, and set $T = T_{\textrm{L}}$, so that equilibrium is recovered in the limit $T_{\textrm{R}} \to T_{\textrm{L}}$.
Nonequilibrium quantities in this paper thus refer to deviations from thermal equilibrium at temperature $T_{\textrm{L}}$. When interpreting the results, however, we occasionally use an
effective local temperature as a convenient expression, as noted below.

\subsection{Equations of motion}

The state of the system is characterized by three variables. These are the real order parameter $\Delta(x,t)$, the nonequilibrium part of the distribution function
$\delta f_{\textrm{L}}(x,t,\omega)$, and a dimensionless complex variable
$\Theta(x,t,\omega)$. The variable $\omega$ denotes the quasiparticle energy, and
$\Theta$ parametrizes the retarded and advanced Green's
functions in the Usadel equation~\cite{usadel1970generalized}.
We decompose the local distribution function for the energy mode as
\begin{equation}
  f_{\textrm{L}}(x,t,\omega) = f^{(0)}_{\textrm{L}}(\omega) + \delta f_{\textrm{L}}(x,t,\omega).
\end{equation}
Here $f^{(0)}_{\textrm{L}}(\omega) \equiv \tanh\!\left(\omega/2k_{\textrm{B}}T\right)$
is the equilibrium distribution, related to the Fermi--Dirac distribution
$n_{\textrm{FD}}(\omega) \equiv \left[\exp\!\left(\omega/k_{\textrm{B}}T\right)+1\right]^{-1}$
by $f^{(0)}_{\textrm{L}}(\omega) = 1 - 2n_{\textrm{FD}}(\omega)$, where $k_{\textrm{B}}$ is the Boltzmann constant. The spectral functions are defined through the real and imaginary parts of
$\cos\Theta$ and $\sin\Theta$:
\begin{subequations}
\begin{align}
\cos\Theta(x,t,\omega) &= N_1(x,t,\omega) + iR_1(x,t,\omega), \\
\sin\Theta(x,t,\omega) &= N_2(x,t,\omega) + iR_2(x,t,\omega).
\end{align}
\end{subequations}
Here $N_1$ represents the local density of states normalized to its normal-state
value at the Fermi level.

These variables evolve according to the following coupled equations~\cite{KramerWattsTobin1978}:
\begin{widetext}
\begin{subequations}
\label{fullEqs}
\begin{align}
  \label{fullRUsadel}
  &\hbar D \frac{\partial^2}{\partial x^2}\Theta + \left(2i\omega-\frac{\hbar}{\tau_{\textrm{E}}}\right)\sin \Theta + 2|\Delta| \cos \Theta=0, \\
  \label{fullTDGL}
  &\left[\gamma_1\frac{\partial}{\partial t}- \frac{1}{|\Delta| \epsilon} \int_0^{\infty} d\omega \; R_2 \, \delta f_{\textrm{L}}\right] \Delta= \xi^2(T)\frac{\partial^2}{\partial x^2} \Delta + \left( 1 - \frac{|\Delta|^2}{\Delta_{\infty}^2(T)} \right) \Delta, \\
  \label{fullBoltzmann}
  &D \frac{\partial }{\partial x} \left[ \left( N_1^2 - R_2^2 \right) \frac{\partial }{\partial x} \delta f_{\textrm{L}} \right] - N_1 \left( \frac{\partial}{\partial t} + \frac{1}{\tau_{\textrm{E}}} \right) \delta f_{\textrm{L}}= R_2 \frac{d f^{(0)}_{\textrm{L}}}{d \omega} \frac{\partial |\Delta|}{\partial t}.
\end{align}
\end{subequations}
\end{widetext}
Equation~\eqref{fullRUsadel} is the Usadel equation generalized to the time-dependent case with inelastic scattering and determines the spectral functions.
Equation~\eqref{fullTDGL} is the generalized TDGL equation, where the second term on the left-hand side describes the coupling between the nonequilibrium distribution and the order parameter. We refer to this contribution as \textit{the anomalous term}, since it is absent in the conventional TDGL equation. We denote it by
\begin{equation} \label{def_Q}
  \mathcal{Q}(x,t) \equiv \frac{1}{|\Delta| \epsilon} \int_0^{\infty} d\omega \; R_2 \, \delta f_{\textrm{L}}.
\end{equation}
Equation~\eqref{fullBoltzmann} is the transport
equation for $\delta f_{\textrm{L}}$, with the right-hand side acting as a
source driven by the time variation of $|\Delta|$.

The parameters appearing in Eqs.~\eqref{fullEqs} are defined as follows: $\hbar$ is the reduced Planck constant, $D$ is the diffusion constant associated with impurity scattering, $\tau_{\textrm{E}}$ is the electron--phonon inelastic scattering time, and $\gamma_1 = \pi\hbar/(8k_{\textrm{B}}T_{\textrm{c}}\epsilon)$ is the TDGL relaxation time. We also define $\epsilon \equiv (T_{\textrm{c}}-T)/T_{\textrm{c}}$,
the coherence length
$\xi(T) = \sqrt{\pi\hbar D/(8k_{\textrm{B}}T_{\textrm{c}}\epsilon)}$,
and the bulk equilibrium order parameter
$\Delta_{\infty}(T) = \sqrt{8\pi^2(k_{\textrm{B}}T_{\textrm{c}})^2\epsilon/[7\zeta(3)]}$.
Here $\zeta(\cdot)$ is the Riemann zeta function ($\zeta(3)\simeq 1.20$).

\subsection{Boundary conditions}

We impose DW-type boundary conditions on $\Delta$~\cite{richard2016heat,Kanakubo2026}:
\begin{subequations} \label{BCsDelta}
\begin{align}
\Delta(-L/2,t) &= \Delta_{\infty}(T_{\textrm{L}})\tanh\left(\frac{-L/2}{\sqrt{2}\,\xi(T_{\textrm{L}})}\right), \label{BCsDeltaleft} \\
\Delta(+L/2,t) &= \Delta_{\infty}(T_{\textrm{R}})\tanh\left(\frac{+L/2}{\sqrt{2}\,\xi(T_{\textrm{R}})}\right). \label{BCsDeltaright}
\end{align}
\end{subequations}
These are motivated by the exact DW solution of the static Ginzburg--Landau
equation for an infinite uniform system at temperature $T$,
\begin{equation}
\xi^2(T)\frac{d^2\Delta}{dx^2}+\Delta-\frac{\Delta^3}{\Delta_{\infty}^2(T)}=0,
\end{equation}
which reads
\begin{equation} \label{tanhprofile}
\Delta(x)=\Delta_{\infty}(T)\tanh\left(\frac{x}{\sqrt{2}\,\xi(T)}\right).
\end{equation}
Equations~\eqref{BCsDeltaleft} and \eqref{BCsDeltaright} fix $\Delta$ at the left and right boundaries to the equilibrium values corresponding to $T_{\textrm{L}}$ and $T_{\textrm{R}}$, respectively. These boundary values impose a DW configuration across the system. We note that $T_{\textrm{R}}$ enters only
through this boundary value. The coefficients $\epsilon$, $\xi(T)$, and
$\Delta_{\infty}(T)$ in Eqs.~\eqref{fullEqs} remain evaluated at $T=T_{\textrm{L}}$.

The bulk solution of Eq.~\eqref{fullRUsadel} for a uniform system at temperature
$T$ is
\begin{equation} \label{0thBCsforTheta}
  \Theta_{\textrm{bulk}} = \arctan\!\left( \frac{-2\Delta_{\infty}(T)}{2i\omega - \hbar/\tau_{\textrm{E}}}\right),
\end{equation}
where the branch is chosen so that $N_1 = \textrm{Re}[\cos\Theta] > 0$, ensuring
the density of states is positive. The boundary conditions for $\Theta$ are
therefore
\begin{subequations} \label{BCsTheta}
\begin{align}
  \Theta(-L/2,t,\omega) &= \arctan\!\left( \frac{-2\Delta_{\infty}(T_{\textrm{L}})}{2i\omega-\hbar/\tau_{\textrm{E}}}\right), \\
  \Theta(+L/2,t,\omega) &= \arctan\!\left( \frac{-2\Delta_{\infty}(T_{\textrm{R}})}{2i\omega-\hbar/\tau_{\textrm{E}}}\right).
\end{align}
\end{subequations}

Finally, the boundary conditions for $\delta f_{\textrm{L}}$ are
\begin{subequations} \label{BCsdeltafL}
\begin{align}
  \delta f_{\textrm{L}}(-L/2,t,\omega) &= 0, \\
  \delta f_{\textrm{L}}(+L/2,t,\omega) &=
  \tanh\!\left(\frac{\omega}{2k_{\textrm{B}}T_{\textrm{R}}}\right)
  -\tanh\!\left(\frac{\omega}{2k_{\textrm{B}}T_{\textrm{L}}}\right).
\end{align}
\end{subequations}
The first condition reflects that quasiparticles at the left boundary are in
thermal equilibrium at $T_{\textrm{L}}$. The second condition states that the
right reservoir at $T_{\textrm{R}} > T_{\textrm{L}}$ injects a nonequilibrium
component. Note that $\delta f_{\textrm{L}}(+L/2,t,\omega) < 0$ for $\omega > 0$.

\section{Numerical solution} \label{sec_num}
We solve the equations of motion~\eqref{fullEqs} numerically under the
boundary conditions~\eqref{BCsDelta}, \eqref{BCsTheta}, and \eqref{BCsdeltafL}.
We choose the numerical parameters
$L/\xi=20$, $T_{\textrm{L}}/T_{\textrm{c}}=0.92$, $T_{\textrm{R}}/T_{\textrm{c}}=0.95$,
$\Delta_{\infty}\xi^2/(\hbar D)=1.0$, and
$\tau_{\textrm{E}}\Delta_{\infty}/\hbar=10.0$,
where $\xi \equiv \xi(T_{\textrm{L}})$ and $\Delta_{\infty} \equiv \Delta_{\infty}(T_{\textrm{L}})$.
We cut off the integration over the quasiparticle energy at
$\omega_{\textrm{max}}=30.0\,\Delta_{\infty}(T_{\textrm{L}})$.
We compute the time evolution as follows. First, we prepare an
initial state at $t=0$ that satisfies the boundary conditions. For a given
$\Delta$, we then solve the time-independent equation~\eqref{fullRUsadel} at
each point on the quasiparticle energy mesh to obtain $N_1$ and $R_2$. We
next integrate Eqs.~\eqref{fullTDGL} and \eqref{fullBoltzmann} forward in
time to update $\Delta$ and $\delta f_{\textrm{L}}$, and repeat this
procedure from the solution of Eq.~\eqref{fullRUsadel}.
We discretize the spatial derivatives by second-order central differences and
integrate in time with an implicit-explicit (IMEX) Runge-Kutta
scheme~\cite{ascher1997implicit, kennedy2003additive, kassam2005fourth}. This scheme treats the diffusive terms implicitly
and the nonlinear terms explicitly. We solve Eq.~\eqref{fullRUsadel} by
Newton's method at each $x$ and $\omega$.
For the initial state at $t=0$, we construct a $\tanh$-type profile for
$\Delta$ that satisfies the boundary conditions~\eqref{BCsDelta} exactly. We
obtain $\delta f_{\textrm{L}}$ by linear interpolation between the
boundary values~\eqref{BCsdeltafL}. The results below show these initial
profiles for each quantity as the $t=0$ snapshot.

\subsection{Solution of $\Delta$}
\begin{figure*}[tb]
\includegraphics[width=153mm]{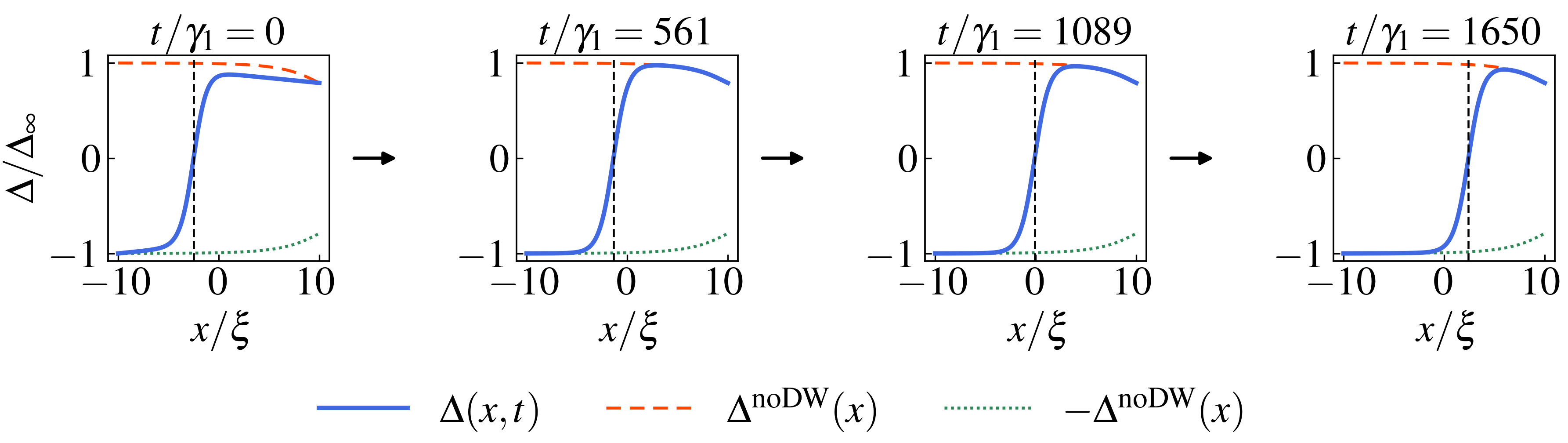}
\caption{Snapshots of the numerical solution of $\Delta$ at four different
times, from left to right. The horizontal axis is the spatial coordinate $x$
scaled by $\xi \equiv \xi(T_{\textrm{L}})$, and the vertical axis is $\Delta$
scaled by $\Delta_{\infty}(T_{\textrm{L}})$. The blue solid curve is the
numerical solution of Eqs.~\eqref{fullEqs} under the boundary
conditions~\eqref{BCsDelta}, \eqref{BCsTheta}, and \eqref{BCsdeltafL}. The
red and green dashed curves are the steady-state solutions
$\Delta^{\textrm{noDW}}(x)$ and $-\Delta^{\textrm{noDW}}(x)$, respectively,
obtained without a domain wall (see text). The black vertical dashed line
marks the domain wall position $x_0$, defined by $\Delta(x_0,t)=0$. The
domain wall moves toward the higher-temperature boundary ($x=+L/2$) as time
increases.}
\label{figDelta}
\end{figure*}

Figure~\ref{figDelta} shows four snapshots of the time evolution of the order
parameter $\Delta$, arranged in chronological order from left to right. The
horizontal axis is the spatial coordinate scaled by $\xi$, and the vertical
axis is the order parameter scaled by $\Delta_{\infty}(T_{\textrm{L}})$. The blue solid curve shows the numerical solution at each time. The red dashed curve shows the steady-state solution $\Delta^{\textrm{noDW}}(x)$
obtained without a DW. It is calculated by solving the equations of
motion~\eqref{fullEqs} with
$\Delta(-L/2,t) = -\Delta_{\infty}(T_{\textrm{L}})\tanh\!\left(-L/2/(\sqrt{2}\,\xi(T_{\textrm{L}}))\right)>0$
in place of the boundary condition~\eqref{BCsDeltaleft}. The green dashed
curve shows $-\Delta^{\textrm{noDW}}(x)$. In contrast to these no-DW
reference solutions, the blue solid curve satisfies the DW boundary
conditions~\eqref{BCsDelta}. Because these boundary conditions impose
opposite signs of $\Delta$ at the two boundaries, the numerical solution
must have a coordinate $x_0$ at which $\Delta(x_0,t)=0$. The black vertical
dashed line marks this domain-wall position $x=x_0$.

The DW moves in the positive direction, that is,
toward the higher-temperature boundary. Far from the DW structure, the
numerical solution asymptotically follows the red or green dashed curve.
Because the boundary conditions force the system to host a DW, the solution
switches from one branch to the other within the interior of the system.

\subsection{Solution of $N_1$ and $R_2$}
\begin{figure*}[tb]
\includegraphics[width=153mm]{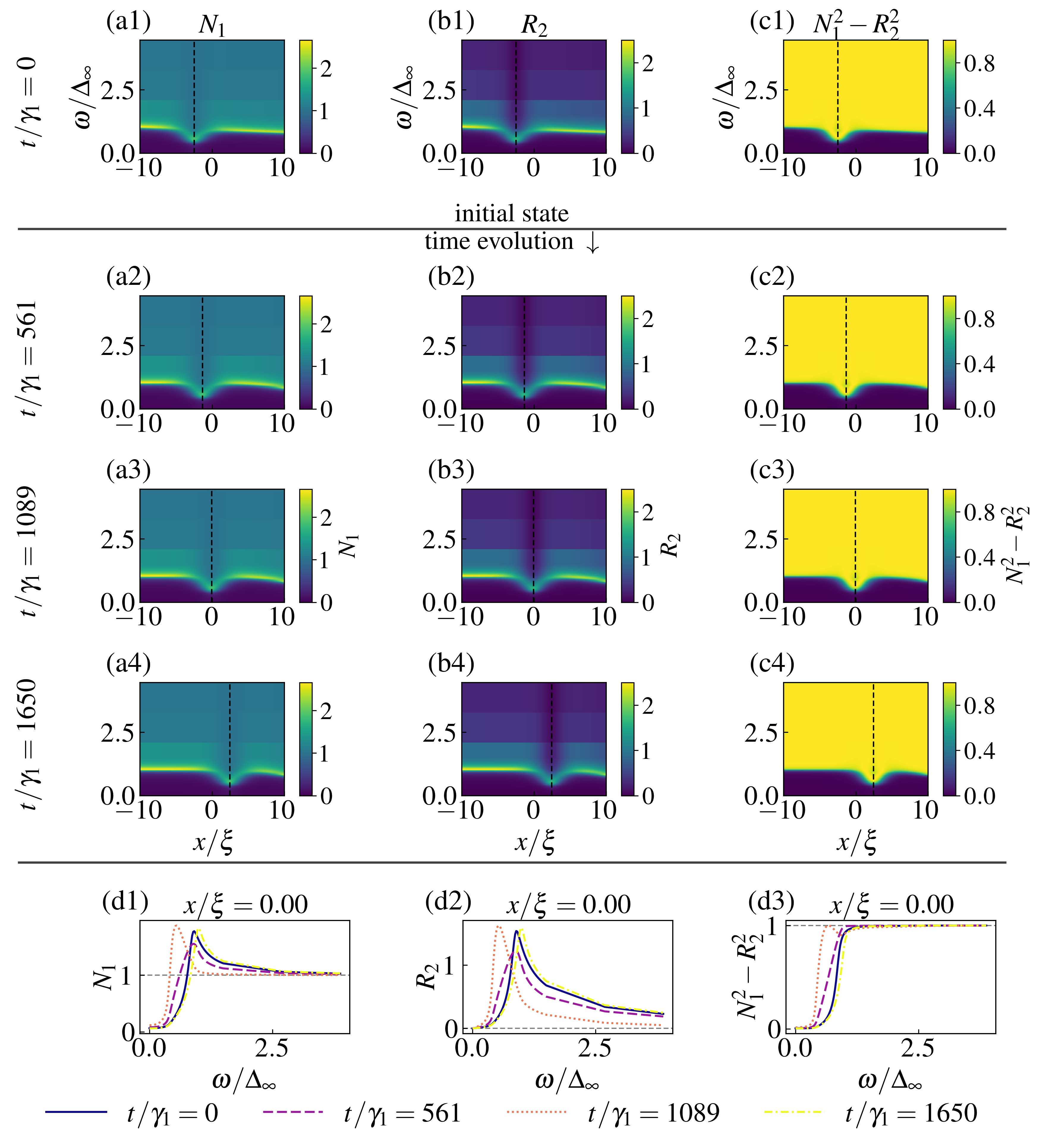}
\caption{Snapshots of the spectral quantities (a) $N_1$, (b) $R_2$, and (c)
$N_1^2-R_2^2$ as functions of $x$ and $\omega$, at four times from top to
bottom. Panels (a1), (b1), (c1) show the initial state. The black vertical
dashed line marks the DW coordinate $x_0(t)$. Row (d) shows the $\omega$
dependence at fixed $x=0$ for the four times shown above.}
\label{figspectral}
\end{figure*}

Figure~\ref{figspectral} shows snapshots of the time evolution of (a) $N_1$,
(b) $R_2$, and (c) $N_1^2-R_2^2$ as functions of $x$ and $\omega$. Panels
(a1), (b1), and (c1) show the initial state. The black vertical dashed line
marks $x=x_0(t)$. Row (d) fixes $x=0$ and shows the $\omega$ dependence at
several times.

Both $N_1$ and $R_2$ exhibit a peak at $\omega \sim
\Delta_{\infty}$ along the energy axis, and approach $N_1
\rightarrow 1$ and $R_2 \rightarrow 0$ for $\omega \gg \Delta_{\infty}$. In particular, row (a) and panel (d1) show that inelastic scattering rounds
off the coherence peak. $N_1$ remains finite as $\omega \rightarrow 0$ at every time shown, and its line shape is similar to that produced by the Dynes phenomenological broadening of the BCS density of states~\cite{dynes1978direct}.

The coefficient $N_1^2-R_2^2$ appearing in the transport equation~\eqref{fullBoltzmann} rises from zero over an energy scale of order $\Delta_{\infty}$ and saturates to $N_1^2-R_2^2 \rightarrow 1$ for $\omega \gg \Delta_{\infty}$.

Turning to the spatial dependence, we focus on the peak structure near
$\omega \sim \Delta_{\infty}$. For both spectral quantities, this peak shifts toward lower $\omega$ in the vicinity of the DW coordinate $x \sim x_0$, tracking the local suppression of $|\Delta|$ there. Reflecting the boundary conditions, it also shifts slightly toward lower $\omega$ overall as $x$ increases. The peak height, by contrast, changes little along $x$.

\subsection{Solution of $\delta f_{\textrm{L}}$}

\begin{figure*}[tb]
\includegraphics[width=163mm]{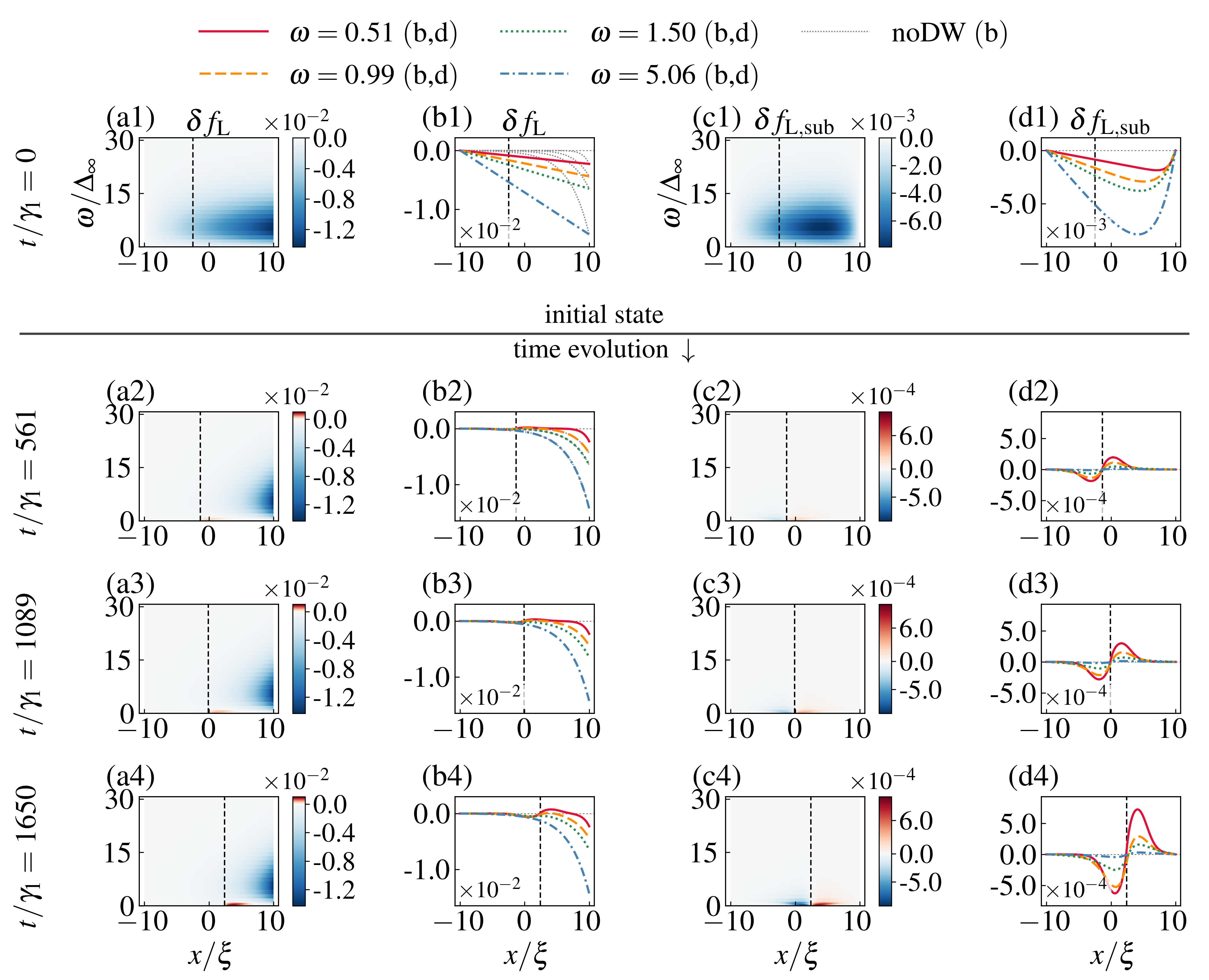}
\caption{Snapshots of the numerical solution of the nonequilibrium
distribution function $\delta f_{\textrm{L}}$ at four times, from top to
bottom. Column (a) shows $\delta f_{\textrm{L}}(x,\omega)$ as a colormap.
Column (b) shows the spatial dependence of $\delta f_{\textrm{L}}$ at four
fixed energies $\omega/\Delta_{\infty}=0.51, 0.99, 1.50, 5.06$. The gray
dotted curves in panel (b1) show the steady-state solution
$\delta f_{\textrm{L}}^{\textrm{noDW}}$ obtained without a DW. Column (c)
shows the colormap of $\delta f_{\textrm{L}}-\delta f_{\textrm{L}}^{\textrm{noDW}}$,
that is, the solution with the noDW background subtracted. Note the
different color scale from column (a). Column (d) shows the spatial
dependence of the subtracted solution at the same four energies as in
column (b). The black vertical dashed line marks the DW coordinate $x_0(t)$,
defined by $\Delta(x_0,t)=0$.}
\label{figdfl}
\end{figure*}

Figure~\ref{figdfl} shows the numerical solution of the modulation of the nonequilibrium distribution function, $\delta f_{\textrm{L}}$. Column (a) shows chronological snapshots of colormaps of $\delta f_{\textrm{L}}$ as a function of $x$ and $\omega$. Column (b) shows the spatial dependence of the solution shown in (a) at the corresponding time, taken at four fixed energies and overlaid in a single plot. The gray dotted curves in
panel (b1) show the steady-state solution $\delta f_{\textrm{L}}^{\textrm{noDW}}$
obtained without a domain wall. For each energy, the solid curve and the corresponding gray dotted curve share the same boundary value and represent the DW and noDW solutions, respectively. Column (c) shows the corresponding colormaps after subtracting the noDW component from the solution in column (a). Note that the color scale differs from that of (a). Column (d) shows the spatial dependence of the solution shown in (c) at the corresponding time, taken at the same four fixed energies as in column (b).

These results show the following. As shown in (c) and (d), the solution
obtained after subtracting the noDW component takes finite values only in a limited region, over an energy range of order $\Delta_{\infty}$ and in the vicinity of $x \sim x_0$. This structure follows the motion of the DW. This indicates that the full solution $\delta f_{\textrm{L}}$ consists of the noDW component as a background, together with the component associated with the DW motion shown in (c) and (d). Section~\ref{sec:efftemp} formalizes this separation as an exact
decomposition of $\delta f_{\textrm{L}}$ into a boundary-driven part and a
DW-motion-induced part.

The exponential-like profile produced by the noDW component reflects the energy injected from the reservoir at the boundary $x=L/2$, which diffuses into the bulk while relaxing toward the phonon bath and produces nonequilibrium excitations throughout the system. This effect is
expected to be monotonic in $x$. Indeed, as shown in (b1), the
noDW component decreases monotonically with $x$. We note here that the noDW solution takes more negative values near the high-temperature boundary $x \sim L/2$. In contrast, the results in columns (c) and (d) show that the spatial
profile of the distribution function localized at the DW is negative for
$x<x_0$ and turns positive for $x>x_0$. That is, the nonequilibrium
distribution function behaves non-monotonically in the vicinity of the DW.
This indicates that this contribution arises from a mechanism distinct from
the simple thermal-diffusive nonequilibrium excitation caused by the
temperature difference between the boundaries.

This non-monotonicity originates from the DW dynamics. The right-hand side of the transport equation~\eqref{fullBoltzmann} is proportional to the time derivative of the magnitude of the order parameter. As the DW moves, the time derivative of
$|\Delta|$ changes sign across $x=x_0$, while $R_2$ does not change sign
(see Fig.~\ref{figspectral}). The right-hand side of
Eq.~\eqref{fullBoltzmann} therefore takes opposite signs on the two sides of
$x=x_0$. Consequently, this right-hand side acts as a source and a sink of
$\delta f_{\textrm{L}}$ on the two sides of the DW, generating the
non-monotonic, locally dipole-like distribution of $\delta f_{\textrm{L}}$
seen in Fig.~\ref{figdfl}.

\subsection{Force balance}
In this subsection we discuss force balance associated with the DW motion on the basis of the momentum balance relation~\cite{KC, Sugai, Kanakubo2026} derived from the TDGL equation~\eqref{fullTDGL}. Multiplying both sides of the TDGL equation~\eqref{fullTDGL} by $\frac{d \Delta}{dx}$ and integrating the right-hand side by parts, we obtain
\begin{equation} \label{local_momentum_balance_relation}
\begin{split}
  &\gamma_1 \frac{\partial \Delta}{\partial t} \frac{\partial \Delta}{\partial x} - \frac{\partial}{\partial x}\left[\frac{\xi^2}{2}\left(\frac{\partial \Delta}{\partial x}\right)^2+\frac{\Delta^2}{2}-\frac{\Delta^4}{4\Delta_{\infty}^2}\right] \\
  &\quad - \frac{\partial \Delta}{\partial x} \frac{\textrm{sgn}(\Delta)}{\epsilon} \int_0^{\infty} d\omega \; R_2 \, \delta f_{\textrm{L}} = 0.
\end{split}
\end{equation}
This is the local momentum balance relation in the present model. It corresponds to the relation appearing in earlier work on single vortex dynamics within the conventional TDGL equation~\cite{KC, Sugai}. The first term contains the time derivative of $\Delta$ and thus represents momentum dissipation associated with TDGL relaxation. We call it the viscous term. The second term on the left-hand side can be interpreted as a hydrodynamic pressure gradient, or equivalently the divergence of the momentum flux tensor. The third term on the left-hand side gives the contribution from the anomalous term. We define
\begin{subequations}
\label{eq:forceball_definitions_x}
\begin{align}
\label{eq:forceball_fvis}
f_{\textrm{vis}}(x,t)
&\equiv
\gamma_1 \frac{\partial\Delta}{\partial t}\frac{\partial\Delta}{\partial x} \\
\label{eq:forceball_P}
P(x,t) &\equiv -\left(\frac{\xi^2}{2}\left(\frac{\partial \Delta}{\partial x}\right)^2+\frac{\Delta^2}{2}-\frac{\Delta^4}{4\Delta_{\infty}^2}\right) \\
\label{eq:forceball_fan}
f_{\textrm{an}}(x,t)
&\equiv
-\frac{\textrm{sgn}(\Delta)}{\epsilon}
\frac{\partial \Delta}{\partial x} \int_0^{\infty} d\omega \; R_2 \, \delta f_{\textrm{L}}
\end{align}
\end{subequations}

so that the local momentum balance relation~\eqref{local_momentum_balance_relation} takes the form

\begin{equation}
  f_{\textrm{vis}}+ \frac{\partial P}{\partial x} + f_{\textrm{an}} = 0.
\end{equation}

Here, the local balance of the force density is expressed as the sum of three contributions with distinct physical origins.

\begin{figure*}[tb]
\includegraphics[width=163mm]{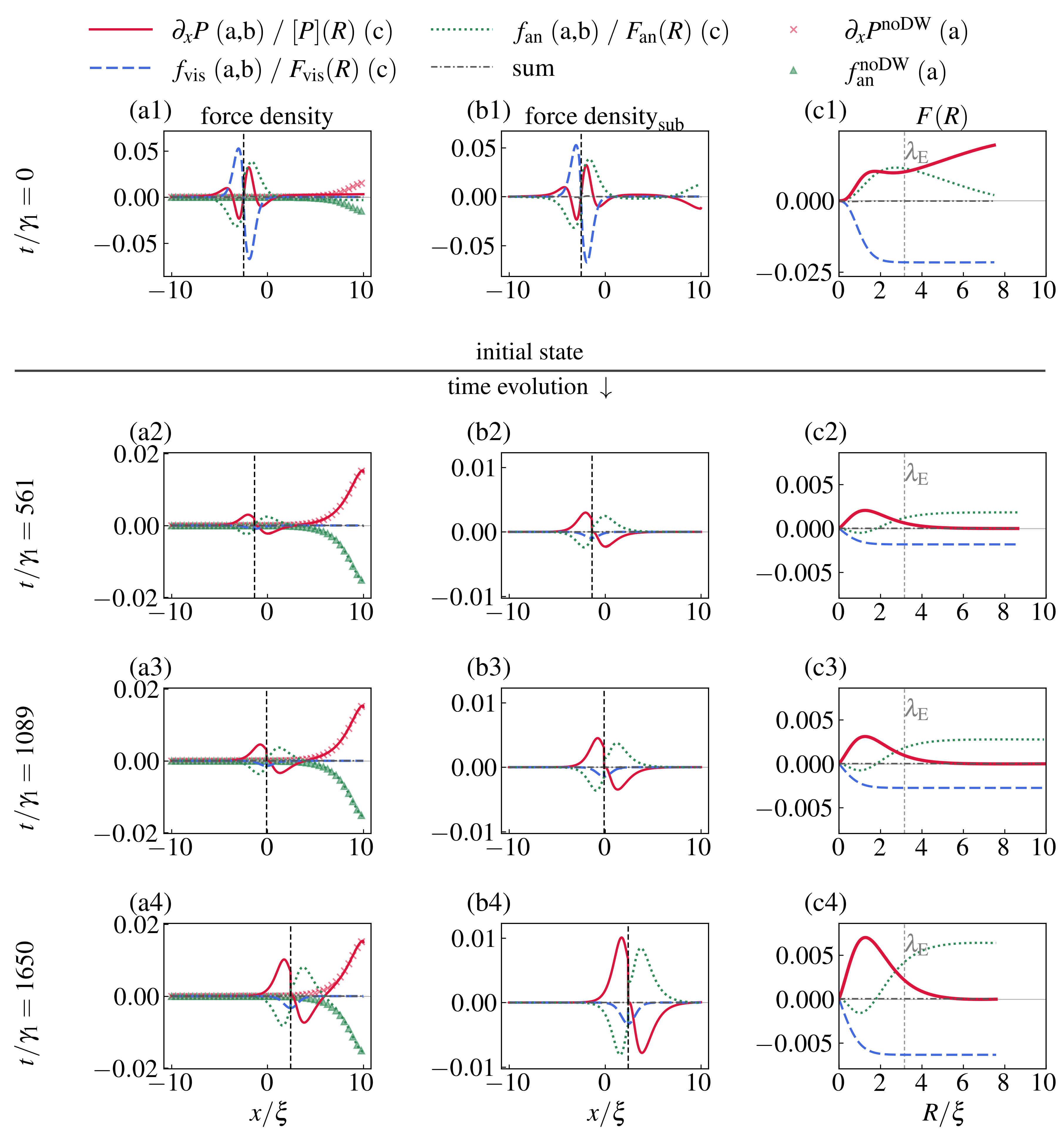}
\caption{Snapshots of the force densities appearing in the local momentum
balance relation~\eqref{local_momentum_balance_relation} at four times, from
top to bottom. Column (a) shows the viscous term $f_{\textrm{vis}}$, the
pressure gradient $\partial P/\partial x$, and the anomalous term
$f_{\textrm{an}}$ as functions of $x$. The markers show the corresponding
steady-state quantities $\partial P^{\textrm{noDW}}/\partial x$ and
$f^{\textrm{noDW}}_{\textrm{an}}$ obtained without a DW, and the gray
dash-dotted curve shows the sum of the three contributions. Column (b) shows
the same quantities after subtracting the noDW contribution ($f_{\textrm{vis}}$
is unchanged, since $\partial\Delta^{\textrm{noDW}}/\partial t=0$). Column (c)
shows the integral $F(R,t)$ of each subtracted force density over the
interval $[x_0(t)-R,\,x_0(t)+R]$, as a function of the half-width $R$. The
vertical gray dashed line marks $\lambda_{\textrm{E}}=\sqrt{D\tau_{\textrm{E}}}$.
The black vertical dashed line in (a) and (b) marks the DW coordinate
$x_0(t)$.}
\label{figForcebalance}
\end{figure*}

Figure~\ref{figForcebalance}(a) shows the spatial dependence of these three force densities as snapshots arranged in time order. Note that (a1) shows the initial state, and the corresponding quantities do not satisfy the dynamical equations~\eqref{fullEqs}. In each panel the black vertical dashed line marks the coordinate $x=x_0(t)$ satisfying $\Delta(x_0,t)=0$, and the horizontal gray dashed line shows the sum of the three contributions. This sum remains on the horizontal axis, confirming that the balance relation is satisfied numerically. The same behavior is also observed in panels (b) and (c). The viscous term, the pressure gradient, and the anomalous term all develop structures at the DW position. The pressure gradient and the anomalous term additionally exhibit variations of comparable magnitude and opposite sign near $x \lesssim L/2$. In this sense $\frac{\partial P}{\partial x}$ and $f_{\textrm{an}}$ are not localized at the DW, whereas the viscous term $f_{\textrm{vis}}$ takes nonzero values only in the vicinity of the DW.

Figure~\ref{figForcebalance}(b) shows the force densities obtained after
subtracting the contribution constructed from the steady-state solution without a domain wall, namely the quantities shown by the markers in panel (a). We denote these steady-state fields by
$\Delta^{\textrm{noDW}}(x)$, $\delta f_{\textrm{L}}^{\textrm{noDW}}(x,\omega)$, $\Theta^{\textrm{noDW}}(x,\omega)$, $N_{1}^{\textrm{noDW}}=\textrm{Re}\left[\cos \Theta^{\textrm{noDW}} \right]$, and $R_{2}^{\textrm{noDW}}=\textrm{Im}\left[\sin \Theta^{\textrm{noDW}} \right]$:

\begin{subequations}
\label{eq:forceball_nodw_definitions}
\begin{align}
  P^{\textrm{noDW}}(x) &\equiv -\left(\frac{\xi^2}{2}\left(\frac{\partial \Delta^{\textrm{noDW}}}{\partial x}\right)^2+\frac{(\Delta^{\textrm{noDW}})^2}{2} \right. \notag \\
  &\qquad \left. {}-\frac{(\Delta^{\textrm{noDW}})^4}{4\Delta_{\infty}^2}\right), \\
  f^{\textrm{noDW}}_{\textrm{an}}(x) &\equiv -\frac{\textrm{sgn}(\Delta^{\textrm{noDW}})}{\epsilon}
\frac{\partial \Delta^{\textrm{noDW}}}{\partial x} \notag \\
&\quad \times \int_0^{\infty} d\omega \; R_{2}^{\textrm{noDW}} \, \delta f_{\textrm{L}}^{\textrm{noDW}}
\end{align}
\end{subequations}

We plot the subtracted quantities $P^{\textrm{sub}}=P-P^{\textrm{noDW}}$ and $f^{\textrm{sub}}_{\textrm{an}}=f_{\textrm{an}}-f^{\textrm{noDW}}_{\textrm{an}}$. For the viscous term the subtracted contribution vanishes because $\frac{\partial \Delta}{\partial t}=0$ for the static solution, so panel (b) shows the same quantity as panel (a). After this subtraction, every component is localized near the DW and moves together with it. We therefore understand panel (b) as panel (a) with the rising structure near $x \lesssim L/2$ in the pressure gradient and the anomalous term removed.

Figure~\ref{figForcebalance}(c) shows $f_{\textrm{vis}}$, $\frac{\partial P^{\textrm{sub}}}{\partial x}$, and $f^{\textrm{sub}}_{\textrm{an}}$ integrated over an interval of half-width $R>0$ centered on the DW coordinate $x_0(t)$ at each time:

\begin{widetext}
\begin{subequations}
\begin{align}
F_{\textrm{vis}}(R,t) &\equiv
\int_{x=x_0(t)-R}^{x=x_0(t)+R} dx\, f_{\textrm{vis}}(x,t), \\
F_{\textrm{an}}(R,t)
&\equiv \int_{x=x_0(t)-R}^{x=x_0(t)+R} dx\, f^{\textrm{sub}}_{\textrm{an}}(x,t), \\
\int_{x=x_0(t)-R}^{x=x_0(t)+R} dx\,\frac{\partial P^{\textrm{sub}}}{\partial x}(x,t)
&=  P^{\textrm{sub}}(x_0(t)+R,t)-P^{\textrm{sub}}(x_0(t)-R,t).
\end{align}
\end{subequations}
\end{widetext}

These quantities represent the force acting on the region of half-width $R$ centered on $x_0$, that is, the region $x_0-R \leq x \leq x_0+R$. The vertical dashed line marks $\lambda_{\textrm{E}}=\sqrt{D\tau_{\textrm{E}}}$, the characteristic length over which the nonequilibrium distribution diffuses before relaxing. Once the integration range becomes sufficiently large, the integrated force in panels (c2)--(c4) no longer depends on the range of integration. This allows us to interpret $f_{\textrm{vis}}$, $\frac{\partial P^{\textrm{sub}}}{\partial x}$, and $f^{\textrm{sub}}_{\textrm{an}}$ as the local force densities acting on the DW. It also lets us define the force acting on the DW itself as the value of the corresponding integral in the limit $R \rightarrow \infty$.

In the limit $R \rightarrow \infty$, panels (c2)--(c4) show that the contribution from the hydrodynamic pressure becomes negligible compared with the contributions from the viscous and anomalous terms. Sufficiently far from the boundaries, the force balance on a DW in the flow state thus reads

\begin{equation}
  F_{\textrm{vis}}(R,t) + F_{\textrm{an}}(R,t) = 0.
\end{equation}

The sign of each force indicates its direction. $F_{\textrm{an}}>0$ drives the DW toward the hot side, whereas $F_{\textrm{vis}}(R,t)<0$ acts as a viscous force opposing the DW motion. This agrees with the picture obtained in Ref.~\cite{Kanakubo2026}, where the DW dynamics is governed by a balance between a thermal force driving the DW toward the hot side and a viscous force opposing the motion.

We now turn to the physical interpretation of the force acting on the DW in the present model. Figure~\ref{figForcebalance}(b) shows that the local density of the anomalous force changes sign across the DW, while its integrated value remains finite. This density is negative for $x<x_0$ and positive for $x>x_0$, indicating that the anomalous force pulls the DW outward on both sides. Quasiparticle excitations suppress the development of the superconducting order parameter. Since these excitations are stronger near the hot boundary, the pull toward the hot side dominates over that toward the cold side, resulting in a net force toward the hot side.

\section{Linear analysis} \label{sec:linear}

We now analyze the DW velocity by a linear-response calculation, the
standard method for treating the flow-state dynamics of vortices and
domain walls~\cite{kopnin1975heat,Dorsey92,KC,Sugai,Kanakubo2026}. Within
this approximation the present model determines the sign of the DW
velocity analytically. The DW moves toward the
region where the superconducting order parameter is suppressed, that is,
toward the higher-temperature boundary.

\subsection{Method}

The linearization procedure follows Ref.~\cite{Kanakubo2026}, where its
validity for DW dynamics was established.

We expand each time-dependent quantity $X = \Theta, \Delta, \delta
f_{\textrm{L}}$ in powers of the DW velocity $v$,
\begin{equation} \label{v_expansion}
  X(x,t) = X^{(0)}(x-vt) + X^{(1)}(x,t) + \mathcal{O}(v^2),
\end{equation}
where the superscripts $0$ and $1$ denote the orders $v^0$ and $v^1$,
respectively. Here $X^{(0)}$ is the equilibrium ($v=0$) profile. We
obtain it as a function of a single spatial argument and evaluate it at
$x-vt$, so that it describes the DW moving rigidly at velocity $v$ to
this order. The correction $X^{(1)}(x,t)$ is not assumed to take the same traveling form. In the linear-response regime, we regard the boundary temperature difference $T_{\textrm{R}}-T_{\textrm{L}}$ and the resulting DW velocity $v$ as quantities of the same order. For the zeroth-order term, the time
derivative reduces to a spatial derivative,
\begin{equation} \label{tw_rule}
  \frac{\partial X^{(0)}(x-vt)}{\partial t} = -v\,\frac{\partial X^{(0)}(x-vt)}{\partial x}.
\end{equation}
We treat the equations for the zeroth- and first-order quantities
separately below.

\subsection{Zeroth-order equations}

At zeroth order, Eqs.~\eqref{fullRUsadel} and \eqref{fullTDGL} become
\begin{equation} \label{0thRUsadel}
\begin{split}
  &\hbar D\,\frac{\partial^2}{\partial x^2}\Theta^{(0)}(x,\omega) +
  \left(2i\omega-\frac{\hbar}{\tau_{\textrm{E}}}\right)\sin
  \Theta^{(0)}(x,\omega) \\
  &\quad + 2|\Delta^{(0)}(x)| \cos
  \Theta^{(0)}(x,\omega)=0,
\end{split}
\end{equation}
\begin{equation} \label{0thTDGL}
  \xi^2 \frac{d^2}{dx^2} \Delta^{(0)}+ \left( 1 -
  \frac{|\Delta^{(0)}|^2}{\Delta_{\infty}^2} \right) \Delta^{(0)} = 0,
\end{equation}
and Eq.~\eqref{fullBoltzmann} is satisfied automatically by
\begin{equation}
  \delta f^{(0)}_{\textrm{L}} = 0.
\end{equation}
This zeroth-order state corresponds to the equilibrium case with no boundary temperature difference. Under the
boundary condition $\Delta^{(0)}(\pm\infty) = \pm\Delta_{\infty}$, the
solution of Eq.~\eqref{0thTDGL} is
\begin{equation} \label{Delta0sol}
  \Delta^{(0)}(x) = \Delta_{\infty} \tanh\!\left(\frac{x}{\sqrt{2}\,\xi}\right),
\end{equation}
consistent with Eq.~\eqref{tanhprofile}. We solve Eq.~\eqref{0thRUsadel}
numerically under the boundary condition
\begin{equation} \label{Theta0BC}
  \Theta^{(0)}(x=\pm L/2, \omega) = \Theta_{\textrm{bulk}}(\omega;T_{\textrm{L}}),
\end{equation}
where $\Theta_{\textrm{bulk}}(\omega;T)$ is the spatially uniform solution
of Eq.~\eqref{fullRUsadel} at temperature $T$, defined analogously to
Eq.~\eqref{0thBCsforTheta} by
\begin{equation} \label{Thetabulkdef}
  \tan \Theta_{\textrm{bulk}}(\omega;T) \equiv
  \frac{-2\Delta_\infty(T)}{2i\omega-\hbar/\tau_{\textrm{E}}}.
\end{equation}
We impose the same value $\Theta_{\textrm{bulk}}(\omega;T_{\textrm{L}})$ at
both boundaries in Eq.~\eqref{Theta0BC}, so that the zeroth-order problem
is closed at the single reference temperature $T_{\textrm{L}}$. We denote
the zeroth-order spectral functions by
\begin{subequations}
\begin{align}
  \cos \Theta^{(0)}(x,\omega) &= N^{(0)}_1(x,\omega) + iR^{(0)}_1(x,\omega), \\
  \sin \Theta^{(0)}(x,\omega) &= N^{(0)}_2(x,\omega) + iR^{(0)}_2(x,\omega).
\end{align}
\end{subequations}
Because $|\Delta^{(0)}(x)|$ is an even function of $x$ and the boundary
values of $\Theta^{(0)}$ coincide at $x=\pm L/2$, both $N_1^{(0)}$ and
$R_2^{(0)}$ are also even functions of $x$. For $\omega>0$, both
$N_1^{(0)}(x,\omega)$ and $R_2^{(0)}(x,\omega)$ are non-negative
throughout the sample. We discuss this in Appendix~\ref{app:R2sign} and
use it without further comment below.

\subsection{First-order equations}

At first order we obtain
\begin{align} \label{1stRUsadel}
 \hbar D\,\frac{\partial^2}{\partial x^2}\Theta^{(1)}(x,\omega)
  + \left(2i\omega - \frac{\hbar}{\tau_{\textrm{E}}}\right)\Theta^{(1)}(x,\omega)\cos \Theta^{(0)} \notag \\
  - 2|\Delta^{(0)}(x)|\Theta^{(1)}(x,\omega)\sin \Theta^{(0)}(x,\omega) \notag \\
  + 2\,\mathrm{sgn}\bigl(\Delta^{(0)}(x)\bigr)\,\Delta^{(1)}(x)\cos \Theta^{(0)} &= 0,
\end{align}
\begin{equation} \label{1stTDGL}
\begin{split}
  &-\gamma_1v\frac{d\Delta^{(0)}}{dx}-\frac{1}{\epsilon \left| \Delta^{(0)}
  \right|}\int_0^{\infty}d\omega \; R^{(0)}_2(x,\omega)\delta
  f^{(1)}_{\textrm{L}}(x,\omega)\Delta^{(0)} \\
  &\quad =\left[
  \xi^2\frac{d^2}{dx^2}+1 - \frac{3\Delta^{(0)2}}{\Delta_{\infty}^2}
  \right]\Delta^{(1)}(x),
\end{split}
\end{equation}
\begin{equation} \label{1stBoltzmann}
\begin{split}
  &D \frac{\partial }{\partial x}
\left[\left(\left(N_1^{(0)}\right)^2
-\left(R_2^{(0)}\right)^2\right)
\frac{\partial }{\partial x}\delta f^{(1)}_{\textrm{L}}(x,\omega)\right] \\
&\quad -\frac{N_1^{(0)}}{\tau_E}\delta f^{(1)}_{\textrm{L}}(x,\omega) \\
  &\quad=-v R_2^{(0)}\frac{d f^{(0)}_{\textrm{L}}(\omega)}{d \omega}\frac{\partial}{\partial x}\left| \Delta^{(0)} \right|.
\end{split}
\end{equation}
We define the linear operator appearing on the right-hand side of
Eq.~\eqref{1stTDGL} by
\begin{equation} \label{def_L_TL}
  \mathcal{L}\,\Delta^{(1)} \equiv
  \xi^2\frac{d^2\Delta^{(1)}}{dx^2}+\left(1-\frac{3\Delta^{(0)2}}{\Delta_\infty^2}\right)\Delta^{(1)},
\end{equation}
and rewrite Eq.~\eqref{1stTDGL} as
\begin{equation} \label{1stTDGL_compact}
\begin{split}
  &-\gamma_1 v\,\frac{d\Delta^{(0)}}{dx} = \mathcal{L}\,\Delta^{(1)} + \mathcal{Q}^{(1)}, \\
  &\mathcal{Q}^{(1)} \equiv \frac{1}{\epsilon}\,\mathrm{sgn}\!\left(\Delta^{(0)}\right)\int_0^\infty d\omega\, R_2^{(0)}\,\delta f_{\textrm{L}}^{(1)},
\end{split}
\end{equation}
where we used $\Delta^{(0)}/|\Delta^{(0)}| = \mathrm{sgn}(\Delta^{(0)})$.
Differentiating Eq.~\eqref{0thTDGL} with respect to $x$ gives
$\mathcal{L}(d\Delta^{(0)}/dx)=0$, so $d\Delta^{(0)}/dx$ is the
translational zero mode of $\mathcal{L}$.

\subsection{Boundary conditions}

We next specify the boundary conditions for these equations. The true, exact time-independent boundary conditions for $\Theta$, $\Delta$, and $\delta f_{\textrm{L}}$ at $x=\pm L/2$ are given by Eqs.~\eqref{BCsDelta},
\eqref{BCsTheta}, and \eqref{BCsdeltafL}.

Under the approximation of Eq.~\eqref{v_expansion}, we must evaluate
$X^{(0)}(x-vt)$ at the fixed boundaries $x=\pm L/2$. Expanding $X^{(0)} =
\Theta^{(0)}, \Delta^{(0)}$ around $x=\pm L/2$ gives
\begin{equation} \label{X0expand}
\begin{split}
  &X^{(0)}(\pm L/2 - vt) \\
  &\quad = X^{(0)}(\pm L/2) - vt\,\frac{dX^{(0)}}{dx}\bigg|_{\pm L/2} + \mathcal{O}(v^2t^2).
\end{split}
\end{equation}
Substituting Eq.~\eqref{X0expand} into the definition $X^{(1)}(\pm
L/2,t) \equiv X(\pm L/2,t) - X^{(0)}(\pm L/2-vt)$ leaves a term
proportional to $vt$ in the boundary value of $X^{(1)}$. However, both
$\Delta^{(0)}$ and $\Theta^{(0)}$ deviate from the DW structure around
$x=0$ only within a region of order $\xi$, so $dX^{(0)}/dx|_{\pm L/2}$ is
exponentially small for $L\gg\xi$. Together with $|vt|\ll\xi$ in the time window considered here, the $vt$ term in Eq.~\eqref{X0expand} is negligible,
and we treat $X^{(1)}(\pm L/2,t)$ as a time-independent constant.

Subtracting the zeroth-order boundary values in Eqs.~\eqref{BCsDelta},
\eqref{BCsTheta}, and \eqref{BCsdeltafL} gives the boundary conditions
for the first-order equations. At the left boundary, all three first-order boundary values vanish, because the zeroth-order problem is defined at the reference temperature $T_{\textrm{L}}$, which coincides with the left boundary temperature. At the right boundary, each first-order boundary value is given by the difference between the corresponding bulk quantity evaluated at $T_{\textrm{R}}$ and at $T_{\textrm{L}}$. These bulk quantities are
$\Theta_{\textrm{bulk}}(\omega;T)$ for $\Theta^{(1)}$,
$\Delta_\infty(T)\tanh\bigl(L/2\sqrt2\,\xi(T)\bigr)$ for $\Delta^{(1)}$,
and $\tanh(\omega/2k_{\textrm{B}}T)$ for $\delta f_{\textrm{L}}^{(1)}$.

Since the linear-response expansion treats $T_{\textrm{R}}-T_{\textrm{L}}$ as the same order as $v$, we keep each of these differences only to leading order in $T_{\textrm{R}}-T_{\textrm{L}}$. To this order, the difference of a bulk quantity between $T_{\textrm{R}}$ and $T_{\textrm{L}}$ reduces to its temperature derivative evaluated at $T_{\textrm{L}}$ multiplied by $T_{\textrm{R}}-T_{\textrm{L}}$.
Differentiating
$\Delta_\infty(T)=k_{\textrm{B}}T_{\textrm{c}}\sqrt{8\pi^2\epsilon(T)/[7\zeta(3)]}$
and $\xi(T)=\sqrt{\pi\hbar D/[8k_{\textrm{B}}T_{\textrm{c}}\epsilon(T)]}$ with
respect to $T$ at fixed $T_{\textrm{c}}$, denoting this derivative by
a prime, and using $d\epsilon/dT=-1/T_{\textrm{c}}$, gives
\begin{equation} \label{DinfxiPrime}
  \Delta_\infty'(T) = -\frac{\Delta_\infty(T)}{2(T_{\textrm{c}}-T)}, \qquad
  \xi'(T) = \frac{\xi(T)}{2(T_{\textrm{c}}-T)}.
\end{equation}
Carrying out this expansion for each of the three bulk quantities above,
and writing $s\equiv 2i\omega-\hbar/\tau_{\textrm{E}}$ and
$u_{\textrm{L}}\equiv L/(2\sqrt2\,\xi(T_{\textrm{L}}))$, gives the
following boundary conditions for the first-order equations,
\begin{subequations} \label{1stBC}
\begin{align}
\Theta^{(1)}(-L/2,\omega)&=0, \label{1stBC_a}\\
\Theta^{(1)}(+L/2,\omega)&=\frac{s\,\Delta_\infty(T_{\textrm{L}})}{\epsilon T_{\textrm{c}}\bigl[s^2+4\Delta_\infty^2(T_{\textrm{L}})\bigr]}(T_{\textrm{R}}-T_{\textrm{L}}), \label{1stBC_b}\\
\Delta^{(1)}(-L/2)&=0, \label{1stBC_c}\\
\Delta^{(1)}(+L/2)&=-\frac{\Delta_\infty(T_{\textrm{L}})}{2\epsilon T_{\textrm{c}}}
\Bigl[\tanh(u_{\textrm{L}})+u_{\textrm{L}}\,\mathrm{sech}^2(u_{\textrm{L}})\Bigr] \notag\\
&\quad \times(T_{\textrm{R}}-T_{\textrm{L}}), \label{1stBC_d}\\
\delta f^{(1)}_{\textrm{L}}(-L/2,\omega)&=0, \label{1stBC_e}\\
\delta f^{(1)}_{\textrm{L}}(+L/2,\omega)&\equiv A(\omega) \notag\\
&=-\frac{\omega}{2k_{\textrm{B}}T_{\textrm{L}}^2}\,\mathrm{sech}^2\!\left(\frac{\omega}{2k_{\textrm{B}}T_{\textrm{L}}}\right)(T_{\textrm{R}}-T_{\textrm{L}}), \label{1stBC_f}
\end{align}
\end{subequations}
where we have dropped terms of $\mathcal{O}\bigl((T_{\textrm{R}}-T_{\textrm{L}})^2\bigr)$
throughout, consistent with the order to which Eqs.~\eqref{1stRUsadel}--\eqref{1stBoltzmann}
are already truncated. Only Eq.~\eqref{1stBC_f} enters the velocity formula
derived in Sec.~\ref{sec:linear}F, through $\delta f_{\textrm{L}}^{\textrm{BC}}$.
Equations~\eqref{1stBC_b} and~\eqref{1stBC_d} are included only to complete the first-order boundary-value problem and are not used in deriving the velocity formula.

\subsection{General solution of the first-order distribution function}

Equation~\eqref{1stBoltzmann} is a linear inhomogeneous equation. We denote by $\varphi_1(x,\omega)$ and $\varphi_2(x,\omega)$ two independent solutions of the corresponding homogeneous equation,
\begin{equation} \label{Boltzmann_hom_TL}
  D \frac{\partial}{\partial x}\!\left[
    \bigl((N_1^{(0)})^2 - (R_2^{(0)})^2\bigr)
    \frac{\partial \varphi}{\partial x}
  \right]
  - \frac{N_1^{(0)}}{\tau_{\textrm{E}}}\,\varphi = 0,
\end{equation}
We also denote by $\Phi_{\textrm{p}}(x,\omega)$ a particular solution of the inhomogeneous equation
\begin{equation} \label{Phip_def}
\begin{split}
  &D \frac{\partial}{\partial x}\!\left[
    \bigl((N_1^{(0)})^2 - (R_2^{(0)})^2\bigr)
    \frac{\partial \Phi_{\textrm{p}}}{\partial x}
  \right]
  - \frac{N_1^{(0)}}{\tau_{\textrm{E}}}\,\Phi_{\textrm{p}} \\
  &\quad =
  -R_2^{(0)}\,\frac{df_{\textrm{L}}^{(0)}}{d\omega}\,\frac{\partial}{\partial x}\left|\Delta^{(0)}\right|.
\end{split}
\end{equation}
The general solution of Eq.~\eqref{1stBoltzmann} is then
\begin{equation} \label{deltaf_decomp_TL}
  \delta f_{\textrm{L}}^{(1)}(x,\omega)
  = c_1(\omega)\,\varphi_1(x,\omega)
  + c_2(\omega)\,\varphi_2(x,\omega)
  + v\,\Phi_{\textrm{p}}(x,\omega).
\end{equation}
Equation~\eqref{Phip_def} does not contain $v$, so $\Phi_{\textrm{p}}$ is
determined by $\Theta^{(0)}$ and $\Delta^{(0)}$ alone. We normalize the homogeneous solutions as
\begin{equation} \label{phi1_norm}
\begin{split}
  &\varphi_1(0,\omega) = 0, \quad \xi\,\left.\frac{\partial\varphi_1}{\partial x}\right|_{x=0} = 1, \\
  &\varphi_2(0,\omega) = 1, \quad \left.\frac{\partial\varphi_2}{\partial x}\right|_{x=0} = 0,
\end{split}
\end{equation}
so that, by the parity of the coefficients in
Eq.~\eqref{Boltzmann_hom_TL}, $\varphi_1$ is an odd function of $x$ and
$\varphi_2$ is an even function of $x$. The coefficients in
Eq.~\eqref{Phip_def} are the same and hence also even in $x$, while its
inhomogeneous term, proportional to $\partial_x|\Delta^{(0)}|$, is odd.
Eq.~\eqref{Phip_def} therefore has an odd solution, and we take $\Phi_{\textrm{p}}$ to be this solution.

We apply the boundary conditions~\eqref{1stBC} to
Eq.~\eqref{deltaf_decomp_TL}. Writing $\varphi_j^R(\omega)\equiv \varphi_j(L/2,\omega)$ ($j=1,2$) and $\Phi_{\textrm{p}}^R(\omega)\equiv \Phi_{\textrm{p}}(L/2,\omega)$ and using the parities of $\varphi_1$, $\varphi_2$, and $\Phi_{\textrm{p}}$, the conditions at $x=\pm L/2$ read
\begin{equation}
\begin{split}
  &c_1\,\varphi_1^R + c_2\,\varphi_2^R + v\,\Phi_{\textrm{p}}^R = A(\omega), \\
  &-c_1\,\varphi_1^R + c_2\,\varphi_2^R - v\,\Phi_{\textrm{p}}^R = 0,
\end{split}
\end{equation}
which give
\begin{equation}
  c_1(\omega) = \frac{A(\omega)-2v\,\Phi_{\textrm{p}}^R(\omega)}{2\,\varphi_1^R(\omega)}, \qquad
  c_2(\omega) = \frac{A(\omega)}{2\,\varphi_2^R(\omega)}.
\end{equation}
Collecting the $v$-dependent part of $c_1$ together with the
$v\,\Phi_{\textrm{p}}$ term, we rewrite Eq.~\eqref{deltaf_decomp_TL} as
\begin{equation} \label{deltaf_BC_Phi_TL}
  \delta f_{\textrm{L}}^{(1)}(x,\omega) = \delta f_{\textrm{L}}^{\textrm{BC}}(x,\omega) + v\,\tilde\Phi_{\textrm{p}}(x,\omega),
\end{equation}
where
\begin{align}
  \delta f_{\textrm{L}}^{\textrm{BC}}(x,\omega) &\equiv
  \frac{A(\omega)}{2\varphi_1^R(\omega)}\varphi_1(x,\omega) +
  \frac{A(\omega)}{2\varphi_2^R(\omega)}\varphi_2(x,\omega), \label{def_deltafBC}\\
  \tilde\Phi_{\textrm{p}}(x,\omega) &\equiv \Phi_{\textrm{p}}(x,\omega) -
  \frac{\Phi_{\textrm{p}}^R(\omega)}{\varphi_1^R(\omega)}\varphi_1(x,\omega). \label{def_tildePhip}
\end{align}
$\delta f_{\textrm{L}}^{\textrm{BC}}$ is independent of $v$ and
represents the diffusive component driven by the boundary temperature
difference. The second term, $v\tilde\Phi_{\textrm{p}}$, represents the component generated by the DW motion itself. $\tilde\Phi_{\textrm{p}}$ is an odd function of $x$ and
vanishes at $x=\pm L/2$.

\subsection{Derivation of the velocity formula}

We multiply both sides of Eq.~\eqref{1stTDGL_compact} by $d\Delta^{(0)}/dx$
and integrate over $x\in(-L/2,L/2)$. The left-hand side becomes
\begin{equation}
  -\gamma_1 v \int_{-L/2}^{L/2}dx\left(\frac{d\Delta^{(0)}}{dx}\right)^{\!2}.
\end{equation}
For the first term on the right-hand side, integration by parts together
with $\mathcal{L}(d\Delta^{(0)}/dx)=0$, and the exponential vanishing of
the boundary terms for $L\gg\xi$, gives
\begin{equation}
  \int_{-L/2}^{L/2}dx\,\frac{d\Delta^{(0)}}{dx}\,\mathcal{L}\Delta^{(1)} = 0.
\end{equation}
For the second term, using $\mathrm{sgn}(\Delta^{(0)})\,d\Delta^{(0)}/dx =
d|\Delta^{(0)}|/dx$, we obtain
\begin{equation}
\begin{split}
  &\int_{-L/2}^{L/2}dx\,\frac{d\Delta^{(0)}}{dx}\,\mathcal{Q}^{(1)} \\
  &\quad = \frac{1}{\epsilon}\int_{-L/2}^{L/2}dx\,\frac{d|\Delta^{(0)}|}{dx}\int_0^\infty d\omega\,R_2^{(0)}\,\delta f_{\textrm{L}}^{(1)}.
\end{split}
\end{equation}
Collecting these results gives
\begin{equation} \label{vselect_TL_pre}
\begin{split}
  &-\gamma_1 v\int_{-L/2}^{L/2}dx\left(\frac{d\Delta^{(0)}}{dx}\right)^{\!2} \\
  &\quad = \frac{1}{\epsilon}\int_{-L/2}^{L/2}dx\,\frac{d|\Delta^{(0)}|}{dx}\int_0^\infty d\omega\,R_2^{(0)}\,\delta f_{\textrm{L}}^{(1)}.
\end{split}
\end{equation}
Substituting Eq.~\eqref{deltaf_BC_Phi_TL} and solving for $v$, we arrive
at
\begin{widetext}
\begin{equation} \label{vselect_TL}
  v = -\frac{\displaystyle\int_{-L/2}^{L/2}dx\,\frac{d|\Delta^{(0)}|}{dx}\int_0^\infty d\omega\,R_2^{(0)}\,\delta f_{\textrm{L}}^{\textrm{BC}}}
  {\displaystyle \epsilon\,\gamma_1\int_{-L/2}^{L/2}dx\left(\frac{d\Delta^{(0)}}{dx}\right)^{\!2} + \int_{-L/2}^{L/2}dx\,\frac{d|\Delta^{(0)}|}{dx}\int_0^\infty d\omega\,R_2^{(0)}\,\tilde\Phi_{\textrm{p}}}.
\end{equation}
\end{widetext}
The numerator represents the diffusive drive imposed by the boundary
temperature difference. In the denominator, the first term is the GL
viscous term and the second is the correction induced by the DW motion
itself. The first term in the denominator is positive. We determine
the sign of the numerator and of the second term in the denominator
below. Section~\ref{sec:efftemp} shows that the same boundary-driven/motion-driven separation also holds for the full transport equation.

We define
\begin{equation} \label{def_calT}
  \mathcal{T}(x) \equiv \int_0^\infty d\omega\,R_2^{(0)}(x,\omega)\,\delta f_{\textrm{L}}^{\textrm{BC}}(x,\omega).
\end{equation}
We subtract the boundary value $|\Delta^{(0)}(L/2)|$ before integrating by
parts, and use $\mathcal{T}(-L/2)=0$, which follows from $\delta
f_{\textrm{L}}^{\textrm{BC}}(-L/2,\omega)=0$. This converts the numerator into
\begin{equation} \label{vselect_TL_altnum}
\begin{split}
  &\int_{-L/2}^{L/2}\!dx\,\frac{d|\Delta^{(0)}|}{dx}\,\mathcal{T}(x) \\
  &\quad = \int_{-L/2}^{L/2}\!dx\,\frac{d\mathcal{T}(x)}{dx}\Bigl(|\Delta^{(0)}(L/2)|-|\Delta^{(0)}(x)|\Bigr).
\end{split}
\end{equation}
Equation~\eqref{vselect_TL} thus becomes
\begin{widetext}
\begin{equation} \label{vselect_TL_alt}
  v = -\frac{\displaystyle\int_{-L/2}^{L/2}\!dx\,\frac{d\mathcal{T}(x)}{dx}\Bigl(|\Delta^{(0)}(L/2)|-|\Delta^{(0)}(x)|\Bigr)}
  {\displaystyle \epsilon\,\gamma_1\int_{-L/2}^{L/2}\!dx\left(\frac{d\Delta^{(0)}}{dx}\right)^{\!2} + \int_{-L/2}^{L/2}\!dx\,\frac{d|\Delta^{(0)}|}{dx}\int_0^\infty\! d\omega\,R_2^{(0)}\,\tilde\Phi_{\textrm{p}}},
\end{equation}
\end{widetext}
which has the same structure as Eq.~(43) of Ref.~\cite{Kanakubo2026}, apart
from $|\Delta^{(0)}|$ appearing here to the first power rather than
squared. This difference traces back to the factor $1/|\Delta|$ in the
definition of $\mathcal{Q}$, Eq.~\eqref{def_Q}, which has no counterpart in
the temperature-linear coefficient of Ref.~\cite{Kanakubo2026}.

The second term in the denominator arises from the motion-induced component $v\tilde\Phi_{\textrm{p}}$ in the decomposition~\eqref{deltaf_BC_Phi_TL}. This is distinct from the boundary-driven part
$\delta f_{\textrm{L}}^{\textrm{BC}}$ in the numerator.
Section~\ref{sec:efftemp} shows that this same contribution renormalizes
the TDGL relaxation coefficient $\gamma_1$. In contrast, the thermal diffusion equation used in Ref.~\cite{Kanakubo2026} contains no such relaxation term. Its heat flow $q$ is thus spatially constant, and the denominator reduces to a single geometric factor. No such conserved flux exists here.

\subsection{Determination of the sign of the velocity}
\label{sec:sign}

We now determine the sign of the velocity in Eq.~\eqref{vselect_TL}.

\subsubsection{Sign of the numerator}

We first examine the parity of the integrand in the numerator.
$\partial_x|\Delta^{(0)}|$ is odd in $x$, while $R_2^{(0)}(x,\omega)$ is even in $x$, as established above. In $\delta
f_{\textrm{L}}^{\textrm{BC}}$, the $\varphi_2$ component is even and the
$\varphi_1$ component is odd. The product $\partial_x|\Delta^{(0)}|\cdot
R_2^{(0)}\cdot\varphi_2$ is odd $\times$ even $\times$ even $=$ odd, so it
vanishes upon integration over $x$. Only the $\varphi_1$ component
survives,
\begin{align} \label{numerator_simplified}
  &\int_{-L/2}^{L/2}\!dx\;\frac{\partial|\Delta^{(0)}|}{\partial x} \int_0^\infty\!d\omega\;R_2^{(0)}\,\delta f_{\textrm{L}}^{\textrm{BC}} \notag \\
  &= \int_{-L/2}^{L/2}\!dx\;\frac{\partial|\Delta^{(0)}|}{\partial x} 
  \int_0^\infty\!d\omega\;R_2^{(0)}\,\frac{A(\omega)}{2\,\varphi_1^R(\omega)}\,\varphi_1(x,\omega).
\end{align}

We next determine the sign of each factor. For $\omega>0$, Eq.~\eqref{1stBC_f} gives $A(\omega) =
-\dfrac{\omega}{2k_{\textrm{B}}T_{\textrm{L}}^2}\,\mathrm{sech}^2\!\left(\dfrac{\omega}{2k_{\textrm{B}}T_{\textrm{L}}}\right)(T_{\textrm{R}}-T_{\textrm{L}}) < 0$ because
$T_{\textrm{R}}>T_{\textrm{L}}$. For $\omega>0$, $R_2^{(0)}(x,\omega)$ is
positive throughout the sample (Appendix~\ref{app:R2sign}). In the BCS limit it is positive for $\omega>|\Delta^{(0)}(x)|$
and vanishes for $\omega<|\Delta^{(0)}(x)|$. At finite $\tau_{\textrm{E}}$
this step is smoothed by the spectral broadening, but its sign is
unchanged. $\partial_x|\Delta^{(0)}|$ is positive for $x>0$ and negative
for $x<0$. Because $\varphi_1$ is odd and, from the
normalization~\eqref{phi1_norm}, satisfies $\partial_x\varphi_1(0,\omega)>0$,
$\varphi_1$ is positive in a neighborhood of $x=0^+$.

We show that $\varphi_1$ remains positive and monotonically increasing
throughout $x>0$, so that in particular $\varphi_1^R>0$. This argument relies on the positivity of $N_1^{(0)}$ and $K^{(0)}(x,\omega)\equiv(N_1^{(0)})^2-(R_2^{(0)})^2$, established in Appendix~\ref{app:R2sign}.
Suppose, for contradiction, that $\varphi_1$ has a local maximum at some
$x_m>0$. At $x_m$, $\partial_x\varphi_1(x_m)=0$, so
Eq.~\eqref{Boltzmann_hom_TL} reduces to
\begin{equation} \label{contradiction}
  DK^{(0)}(x_m)\,
  \frac{\partial^2\varphi_1}{\partial x^2}\bigg|_{x_m}
  = \frac{N_1^{(0)}(x_m)}{\tau_{\textrm{E}}}\,\varphi_1(x_m).
\end{equation}
The right-hand side is positive because $N_1^{(0)}>0$, $\tau_{\textrm{E}}>0$, and $\varphi_1(x_m)>0$, since $\varphi_1$ has already become positive for $x>0$.
Since $K^{(0)}(x_m)>0$, the left-hand side then requires
$\partial_x^2\varphi_1(x_m)>0$, which contradicts the condition for a
maximum, $\partial_x^2\varphi_1(x_m)\leq0$. Hence $\varphi_1$ has no local
maximum for $x>0$. Since $\varphi_1(0)=0$ and $\partial_x\varphi_1(0)>0$,
$\varphi_1$ starts positive just to the right of $x=0$ and remains positive and monotonically increasing throughout $x>0$.
In particular, $\varphi_1^R>0$.

Combining these signs, the integrand of Eq.~\eqref{numerator_simplified} is negative on both sides of the DW. For $x>0$ it has the sign $(+)(+)(-)/[(+)(+)] < 0$, and for $x<0$ it has the sign $(-)(+)(-)/[(+)(-)] < 0$. The quantity in Eq.~\eqref{numerator_simplified} is therefore negative.

\subsubsection{Sign of the second term in the denominator}

We show that the second term in the denominator of Eq.~\eqref{vselect_TL}
is non-negative for all $\tau_{\textrm{E}}>0$. Using the definition of
$\tilde\Phi_{\textrm{p}}$, Eq.~\eqref{def_tildePhip}, and its governing
equation~\eqref{Phip_def}, we write this term as the bilinear form
\begin{equation} \label{IDW_main}
\begin{aligned}
  I_{\textrm{DW}} &\equiv \int_{-L/2}^{L/2}dx\, \frac{\partial|\Delta^{(0)}|}{\partial x} \int_0^\infty d\omega\, R_2^{(0)}\,\tilde\Phi_{\textrm{p}} \notag \\
  &= -\int_0^\infty d\omega\, \frac{df_{\textrm{L}}^{(0)}}{d\omega} \int_{-L/2}^{L/2}\!\!dx \notag \\
  &\quad \int_{-L/2}^{L/2}\!\!dx'\, h(x,\omega)\, G(x,x';\omega)\, h(x',\omega).
\end{aligned}
\end{equation}
with $h(x,\omega)\equiv R_2^{(0)}(x,\omega)\,\partial_x|\Delta^{(0)}(x)|$.
Here $G(x,x';\omega)$ is the Green's function of the operator
\begin{equation} \label{def_Ltau}
\begin{split}
  &  \hat{\mathcal{L}}_\tau \equiv D\frac{\partial}{\partial x}\!\left[K^{(0)}\frac{\partial}{\partial x}\right] - \frac{N_1^{(0)}}{\tau_{\textrm{E}}}, \\
  &K^{(0)}(x,\omega) \equiv (N_1^{(0)})^2-(R_2^{(0)})^2,
\end{split}
\end{equation}
defined under the Dirichlet condition $G(\pm L/2,x';\omega)=0$. Because
$K^{(0)}\geq0$ and $N_1^{(0)}\geq0$, $ \hat{\mathcal{L}}_\tau$ is negative
definite, and its Green's function satisfies the quadratic-form
inequality
\begin{equation} \label{GFF_neg_main}
  \int_{-L/2}^{L/2}dx\int_{-L/2}^{L/2}dx'\,f(x)\,G(x,x';\omega)\,f(x') \leq 0
\end{equation}
for any function $f(x)$. The full construction of $G$ and the proof of
Eq.~\eqref{GFF_neg_main} are given in
Appendix~\ref{app:GF_sign}. Since $df_{\textrm{L}}^{(0)}/d\omega =
(2k_{\textrm{B}}T)^{-1}\mathrm{sech}^2(\omega/2k_{\textrm{B}}T)>0$, Eq.~\eqref{GFF_neg_main} applied
to Eq.~\eqref{IDW_main} gives
\begin{equation} \label{IDW_pos_main}
  I_{\textrm{DW}} \geq 0
\end{equation}
for all $\tau_{\textrm{E}}>0$.

\subsubsection{Sign of the velocity}

The denominator of Eq.~\eqref{vselect_TL} is the sum of the strictly
positive GL viscous term $\epsilon\gamma_1\int
dx\,(d\Delta^{(0)}/dx)^2$ and the non-negative correction
$I_{\textrm{DW}}\geq0$. The denominator is therefore always positive. Combined with the negative sign of the numerator obtained above, this shows that $v>0$ for every $\tau_{\textrm{E}}>0$, provided that $R_2^{(0)}(x,\omega)$ remains positive throughout the sample. Appendix~\ref{app:R2sign} proves this at the boundaries and verifies it numerically in the interior, over the full parameter range studied in this work. The DW therefore moves toward the higher-temperature boundary within the linear-response regime of our model. We find no reversal of the direction of motion in any case examined here.

\subsection{Consistency with numerical result}
\begin{figure}[t!]
\includegraphics[width=\columnwidth]{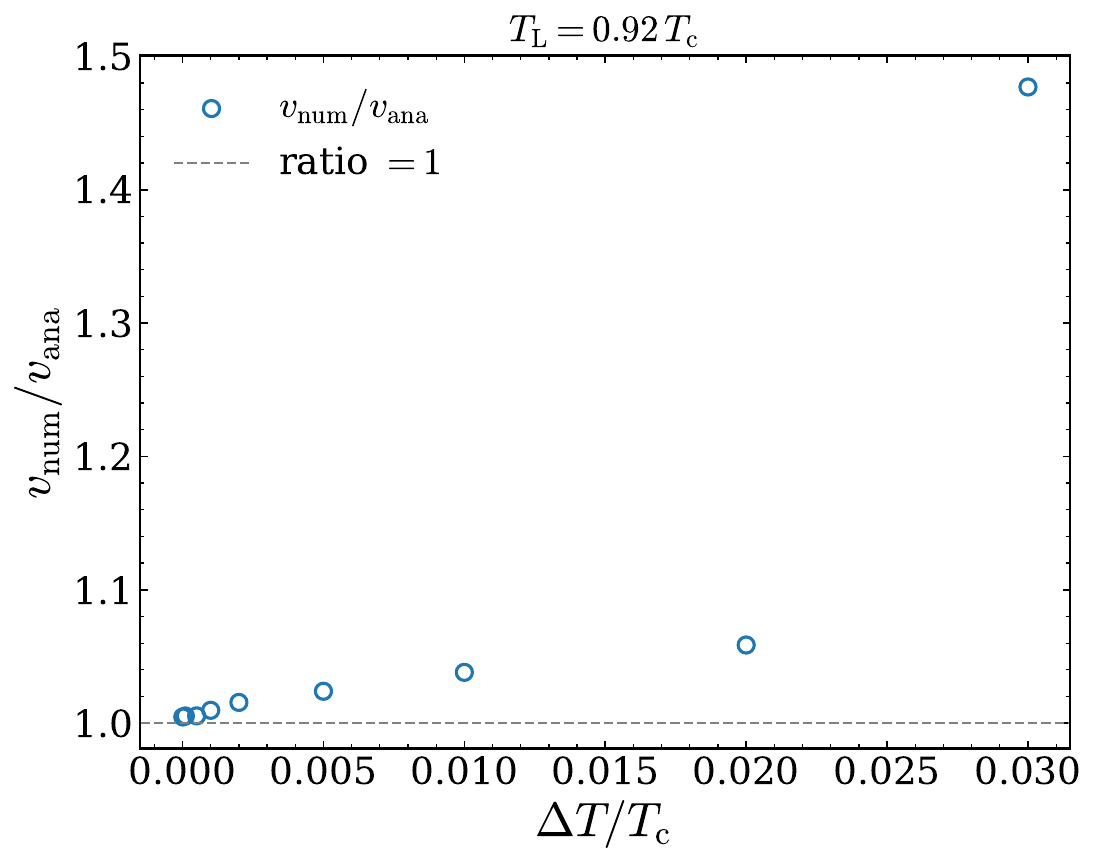}
\caption{Ratio of the DW velocity obtained from the numerical simulation, $v_{\textrm{num}}$, to that obtained from the linear analysis, $v_{\textrm{ana}}$ (Eq.~\eqref{vselect_TL}), as a function of $\Delta T/T_{\textrm{c}}=(T_{\textrm{R}}-T_{\textrm{L}})/T_{\textrm{c}}$, with $T_{\textrm{L}}=0.92\,T_{\textrm{c}}$ fixed. The ratio approaches unity as $\Delta T\to0$, showing that the linear analysis becomes a good approximation for small temperature differences.}
\label{TRsweep}
\end{figure}

The results of the linear analysis presented in this section are also quantitatively consistent with those obtained from numerical simulations. Figure~\ref{TRsweep} shows the ratio of the velocity measured in the numerical simulations to the velocity predicted by the linear analysis, as a function of $\Delta T/T_{\textrm{c}}=(T_{\textrm{R}}-T_{\textrm{L}})/T_{\textrm{c}}$, where $v_{\textrm{num}}$ is obtained from the numerical simulations and $v_{\textrm{ana}}$ from Eq.~\eqref{vselect_TL}. For all data points, $T_{\textrm{L}}=0.92 T_{\textrm{c}}$ is fixed, while $T_{\textrm{R}}$ is varied. We define $v_{\textrm{num}} \equiv \left.\dfrac{d x_0}{dt}\right|_{t=t_0}$, evaluated at the time $t_0$ satisfying $x_0(t_0)=0$. When $\Delta T$ is sufficiently small, the ratio $v_{\textrm{num}}/v_{\textrm{ana}}$ approaches unity, confirming that the linear analysis correctly predicts the DW velocity in this limit. The parameter used in Sec.~\ref{sec_num}, however, corresponds to $\Delta T=0.03$, for which this ratio is about $1.48$. At this larger temperature difference, the linear analysis should thus be regarded as predicting the sign and order of magnitude of the velocity, rather than its precise value.

\section{Discussion} \label{sec:discussion}
This section discusses several aspects of the results presented in the preceding sections.
\subsection{Direction of DW motion}
\label{sec:direction}
Both the numerical calculation and the linear analysis show that the DW moves toward the higher-temperature side even when the anomalous term is included in the TDGL model. The local momentum-balance analysis in Sec.~\ref{sec:sign} traces this direction to the sign of the numerator in the velocity formula, Eq.~\eqref{vselect_TL}. The boundary-driven component $\delta f_{\textrm{L}}^{\textrm{BC}}$ fixes this sign on its own. It vanishes at the cold boundary and grows in magnitude toward the hot boundary, and Sec.~\ref{sec:sign} shows that this growth drives the DW toward the hot side for every $\tau_{\textrm{E}}>0$. The DW-motion-induced component $\tilde\Phi_{\textrm{p}}$ plays a different role. It enters only the denominator, where it adds a non-negative correction $I_{\textrm{DW}}\geq0$ to the GL viscous term. This correction can slow the DW down, but it cannot reverse the direction that the numerator has already fixed. The total $\delta f_{\textrm{L}}$ observed near the DW in Sec.~\ref{sec_num} is the sum of both components, and its non-monotonic profile blends their two contributions together. Thus, the sign of $\delta f_{\textrm{L}}$ at a given point does not by itself determine the direction of motion. The direction is instead fixed by the boundary-driven component $\delta f_{\textrm{L}}^{\textrm{BC}}$.

Earlier work using a TDGL model with an explicit local temperature~\cite{Kanakubo2026} interpreted this direction of motion in terms of condensation energy. A defect such as the DW locally suppresses $|\Delta|$, which costs condensation energy, and this cost is smaller where $|\Delta|$ is already reduced. An impurity suppresses $|\Delta|$ in the same way, which is why it attracts and pins a vortex. Only the boundary-driven component of the anomalous term directly admits this energetic interpretation. It vanishes at the cold boundary, where $|\Delta|$ is largest, and it grows toward the hot boundary, where the boundary-driven anomalous contribution further suppresses $|\Delta|$. It therefore lowers the cost of hosting the DW near the hot boundary. The DW-motion-induced component has no such interpretation, since it does not track a boundary-imposed suppression of $|\Delta|$. As shown above, it instead renormalizes the DW's relaxation through its role in the denominator.

\subsection{Effective temperature}
\label{sec:efftemp}
\subsubsection{Decomposition of the anomalous term}
A description in terms of a local effective temperature would connect our results to Ref.~\cite{Kanakubo2026} and is also convenient for comparison with experiments that report a temperature rather than a nonequilibrium distribution function. We construct such a description below not because it is required to solve our model, but as a diagnostic. It lets us ask what a local-temperature picture would predict once it is imposed on the microscopic distribution function that our model actually uses.

Equation~\eqref{fullBoltzmann} is linear in $\delta f_{\textrm{L}}$, so its solution decomposes into a homogeneous part and a particular part, just as in the linearization used to obtain the DW velocity,
\begin{equation}
  \delta f_{\textrm{L}} = \delta f_{\textrm{L}}^{\textrm{hom}}+\delta f_{\textrm{L}}^{\textrm{part}}.
\end{equation}
The homogeneous part $\delta f_{\textrm{L}}^{\textrm{hom}}$ is driven by the boundary condition on $\delta f_{\textrm{L}}$, that is, by the temperature difference between the two ends, and is expected to represent heat diffusion. The particular part $\delta f_{\textrm{L}}^{\textrm{part}}$, on the other hand, is generated by the DW dynamics through the time-derivative term of $|\Delta|$ on the right-hand side of Eq.~\eqref{fullBoltzmann}. In the rest of this section we discuss $\delta f_{\textrm{L}}$ in terms of these two parts.

The equations of motion and boundary conditions obeyed by each part are
\begin{subequations} \label{eq:delfLhom}
\begin{align}
&D \frac{\partial}{\partial x} \left[ \left( N_1^2 - R_2^2 \right) \frac{\partial}{\partial x} \delta f_{\textrm{L}}^{\textrm{hom}} \right] \notag \\
&\quad {}- N_1 \left( \frac{\partial}{\partial t} + \frac{1}{\tau_{\textrm{E}}} \right) \delta f_{\textrm{L}}^{\textrm{hom}} =0, \label{eq:delfLhom_a} \\
&\delta f_{\textrm{L}}^{\textrm{hom}}(-L/2,t,\omega) = 0, \label{eq:delfLhom_b} \\
&\delta f_{\textrm{L}}^{\textrm{hom}}(+L/2,t,\omega) = \tanh\!\left(\frac{\omega}{2k_{\textrm{B}}T_{\textrm{R}}}\right)
  -\tanh\!\left(\frac{\omega}{2k_{\textrm{B}}T_{\textrm{L}}}\right)
\label{eq:delfLhom_c}
\end{align}
\end{subequations}
and
\begin{subequations} \label{eq:delfLpart}
\begin{align}
&D \frac{\partial}{\partial x} \left[ \left( N_1^2 - R_2^2 \right) \frac{\partial}{\partial x} \delta f_{\textrm{L}}^{\textrm{part}} \right] - N_1 \left( \frac{\partial}{\partial t} + \frac{1}{\tau_{\textrm{E}}} \right) \delta f_{\textrm{L}}^{\textrm{part}} \notag \\
&\quad = R_2 \frac{d f^{(0)}_{\textrm{L}}(\omega)}{d \omega} \frac{\partial |\Delta|}{\partial t}, \label{eq:delfLpart_a} \\
&\delta f_{\textrm{L}}^{\textrm{part}}(\pm L/2,t,\omega) = 0. \label{eq:delfLpart_b}
\end{align}
\end{subequations}
respectively. At first order in $v$, this decomposition reduces to the one used in
Sec.~\ref{sec:linear} to derive the DW velocity, namely
$\delta f_{\textrm{L}}^{\textrm{hom}} \to \delta f_{\textrm{L}}^{\textrm{BC}}$ and
$\delta f_{\textrm{L}}^{\textrm{part}} \to v\tilde\Phi_{\textrm{p}}$, as in
Eq.~\eqref{deltaf_BC_Phi_TL}.

In the limit $T_{\textrm{R}} \rightarrow T_{\textrm{L}}$, the steady homogeneous solution satisfying these boundary conditions vanishes identically. By contrast, the contribution from the particular solution can survive even without an asymmetric boundary condition and is thus not fundamentally tied to the boundary temperature difference. This is consistent with the structure of the KWT equation~\cite{KramerWattsTobin1978}, where the term originating from the particular solution appears together with the time derivative of the order parameter rather than in the coefficient linear in $\Delta$. Motivated by this observation, we split the anomalous term in the TDGL equation into a homogeneous and a particular contribution. This decomposition connects our model to a TDGL description with a local effective temperature.

Using the anomalous term $\mathcal{Q}$ defined in Eq.~\eqref{def_Q}, we write the TDGL equation~\eqref{fullTDGL} as
\begin{align}
&\left[\gamma_1\frac{\partial}{\partial t}- \mathcal{Q}\right] \Delta \notag \\
&\quad = \xi^2\frac{\partial^2}{\partial x^2} \Delta + \left( 1 - \frac{|\Delta|^2}{\Delta_{\infty}^2} \right) \Delta.
\end{align}
The decomposition $\delta f_{\textrm{L}} = \delta f_{\textrm{L}}^{\textrm{hom}}+\delta f_{\textrm{L}}^{\textrm{part}}$ carries directly over to $\mathcal{Q}$:
\begin{subequations} \label{eq:Q_decomp} \begin{align} \mathcal{Q} &= \mathcal{Q}^{\textrm{hom}} + \mathcal{Q}^{\textrm{part}}, \label{eq:Q_decomp_a} \\ \mathcal{Q}^{\textrm{hom}} &= \frac{1}{\epsilon |\Delta|}\int_0^\infty d\omega\, R_2\,\delta f_{\textrm{L}}^{\textrm{hom}}, \label{eq:Q_decomp_b} \\ \mathcal{Q}^{\textrm{part}} &= \frac{1}{\epsilon |\Delta|}\int_0^\infty d\omega\, R_2\,\delta f_{\textrm{L}}^{\textrm{part}}. \label{eq:Q_decomp_c} \end{align} \end{subequations}
The TDGL equation is then rewritten as
\begin{align} \label{tdglhompartdecop}
&\left[\gamma_1\frac{\partial}{\partial t}-\mathcal{Q}^{\textrm{part}}\right] \Delta \notag \\
&\quad = \xi^2\frac{\partial^2}{\partial x^2} \Delta + \left( 1 + \mathcal{Q}^{\textrm{hom}} - \frac{|\Delta|^2}{\Delta_{\infty}^2} \right) \Delta.
\end{align}

The term $\mathcal{Q}^{\textrm{hom}}$ on the right-hand side renormalizes the coefficient that is unity in the equilibrium TDGL equation. Since this coefficient represents the distance from the transition temperature in the GL expansion, its spatial variation can be formally interpreted as a variation of a local effective temperature. This provides a formal correspondence between the present microscopic description and a phenomenological local-temperature description. Solving the equations of our model, however, requires neither this interpretation nor the introduction of an effective temperature.
\subsubsection{$ \delta f_{\textrm{L}}^{\textrm{hom}}$ contribution}

By comparing Eq.~\eqref{tdglhompartdecop} with the dimensionless TDGL equation used in Ref.~\cite{Kanakubo2026}, we define a local effective temperature $T_{\textrm{eff}}(x,t)$ through
\begin{equation}
  1+\mathcal{Q}^{\textrm{hom}} = -\frac{T_{\textrm{eff}}(x,t)-{T_{\textrm{c}}}}{T_{\textrm{c}}-T_{\textrm{L}}},
\end{equation}
so that $\mathcal{Q}^{\textrm{hom}}$ vanishes in the limit $T_{\textrm{eff}} \rightarrow T_{\textrm{L}}$. Solving for $T_{\textrm{eff}}$ and using $\epsilon=(T_{\textrm{c}}-T_{\textrm{L}})/T_{\textrm{c}}$ gives
\begin{equation} \label{def_Teff}
  T_{\textrm{eff}} = T_{\textrm{L}} -\frac{ T_{\textrm{c}}}{|\Delta|}\int_0^\infty d\omega\, R_2\,\delta f_{\textrm{L}}^{\textrm{hom}}.
\end{equation}
The second term represents the deviation from the reference temperature $T_{\textrm{L}}$. Because of the factor $1/|\Delta|$, this term diverges at the DW center. This divergence is not a singularity of the underlying dynamics. It is an artifact of dividing $\mathcal{Q}^{\textrm{hom}}$ by $|\Delta|$ to convert it into a temperature-like quantity. The quantity that actually enters the TDGL equation is the product $\mathcal{Q}^{\textrm{hom}}\Delta = \mathrm{sgn}(\Delta)\,\epsilon^{-1}\int_0^\infty d\omega\,R_2\,\delta f_{\textrm{L}}^{\textrm{hom}}$. This product remains finite through the DW center and only changes sign there, consistent with the finite, sign-changing anomalous force density shown in Fig.~\ref{figForcebalance}.

\subsubsection{Evaluation of the effective temperature}
\begin{figure}[t!]
\includegraphics[width=\columnwidth]{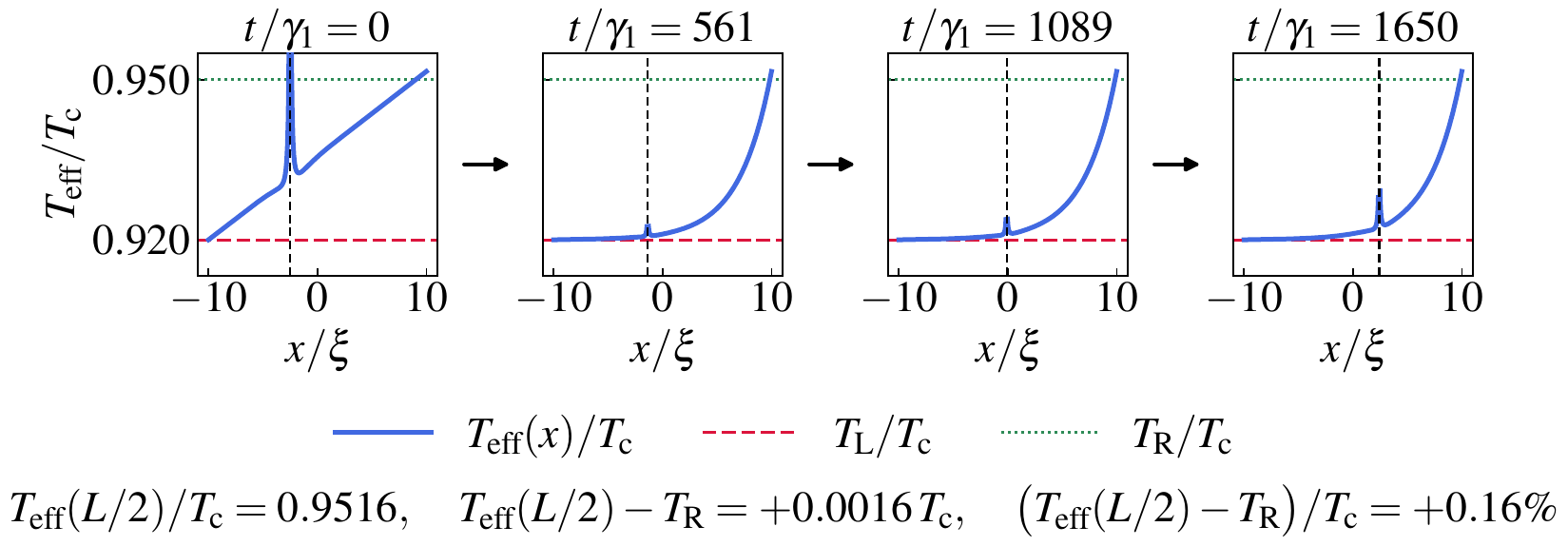}
\caption{Snapshots of the effective local temperature $T_{\textrm{eff}}(x,t)/T_{\textrm{c}}$, obtained from the numerical solution through the definition~\eqref{def_Teff}. The parameters are the same as those used in Sec.~\ref{sec_num}. The black dashed vertical line in each panel marks $x=x_0(t)$. The horizontal guide lines indicate $T_{\textrm{L}}$ and $T_{\textrm{R}}$.}
\label{fig_Teff}
\end{figure}

Figure~\ref{fig_Teff} shows snapshots of the effective temperature. It diverges at $x=x_0(t)$, so this point should not be interpreted physically. Thus, this phenomenological quantity should be used only to interpret the global behavior of the field. Away from the DW center, the effective temperature follows the spatial profile of the nonequilibrium distribution function and has an exponential-like form.

The boundary condition and the definition together require $T_{\textrm{eff}}/T_{\textrm{c}}=T_{\textrm{L}}/T_{\textrm{c}}=0.9200$ at $x=-L/2$. At the boundary $x=L/2$, which is in contact with a bath at temperature $T_{\textrm{R}}$, there is no reason for $T_{\textrm{eff}}$ to coincide with $T_{\textrm{R}}$. $T_{\textrm{eff}}$ is defined by Eq.~\eqref{def_Teff} as a convenient rescaling of $\mathcal{Q}^{\textrm{hom}}$. It is not an independently defined thermodynamic temperature. Evaluating the definition~\eqref{def_Teff} gives $T_{\textrm{eff}}(L/2)/T_{\textrm{c}}=0.9516>T_{\textrm{R}}/T_{\textrm{c}}=0.9500$. This deviation of about $0.16\%$ of $T_{\textrm{c}}$ is not an error. It only reflects the absence of an independent thermodynamic definition for $T_{\textrm{eff}}$. $T_{\textrm{eff}}$ tracks a temperature-like disturbance only where $|\Delta|$ stays away from zero, so this representation necessarily breaks down at the DW center. The physical quantity $\mathcal{Q}^{\textrm{hom}}\Delta$ remains finite there, as noted above.

\subsubsection{$ \delta f_{\textrm{L}}^{\textrm{part}}$ contribution}
Previous work~\cite{KramerWattsTobin1978} evaluated the transport equation approximately in the limit $\lambda_{\textrm{E}}=\sqrt{D\tau_{\textrm{E}}} \ll \xi$, in which the collision term dominates the left-hand side of Eq.~\eqref{eq:delfLpart}, giving
\begin{equation}
  \delta f_{\textrm{L}}^{\textrm{part}} \simeq \frac{-R_2\tau_{\textrm{E}}}{N_1} \frac{d f^{(0)}_{\textrm{L}}(\omega)}{d \omega} \frac{\partial |\Delta|}{\partial t}.
\end{equation}
This algebraic form for $\delta f_{\textrm{L}}^{\textrm{part}}$ is local and instantaneous in $\partial|\Delta|/\partial t$, and it holds only in this limit. Away from this limit, $\delta f_{\textrm{L}}^{\textrm{part}}$ is a nonlocal, retarded functional of $|\Delta|$ determined by Eq.~\eqref{eq:delfLpart}. Substituting this approximate form into the left-hand side of Eq.~\eqref{tdglhompartdecop} gives
\begin{align}
  -\mathcal{Q}^{\textrm{part}}\Delta &= -\frac{1}{\epsilon \left|\Delta\right|} \int_0^\infty d\omega\, \frac{\left(-R_2^2\right)}{N_1}\tau_{\textrm{E}} \frac{df^{(0)}_{\textrm{L}}(\omega)}{d \omega} \frac{\partial \left|\Delta\right|}{\partial t} \Delta \notag \\
  &= \frac{\tau_{\textrm{E}} }{\epsilon} \frac{\partial \Delta}{\partial t}\int_0^\infty d\omega\, \frac{R_2^2}{N_1} \frac{df^{(0)}_{\textrm{L}}(\omega)}{d \omega},
\end{align}
which is proportional to $\partial \Delta/\partial t$. In this limit, the left-hand side of the TDGL equation becomes
\begin{align}
&\left[\gamma_1\frac{\partial}{\partial t}-\mathcal{Q}^{\textrm{part}}\right] \Delta \notag \\
&\quad =  \left(\gamma_1 + \frac{\tau_{\textrm{E}} }{\epsilon} \int_0^\infty d\omega\, \frac{R_2^2}{N_1} \frac{df^{(0)}_{\textrm{L}}(\omega)}{d \omega}\right)\frac{\partial \Delta}{\partial t}.
\end{align}
The second term in this coefficient can be interpreted as a position- and time-dependent correction to $\gamma_1$. Since it is always positive, it increases the effective TDGL relaxation time. This is the short-$\lambda_{\textrm{E}}$ limit of the same non-negative correction discussed in Sec.~\ref{sec:direction}. The exact version enters the denominator of Eq.~\eqref{vselect_TL_alt}.

This same structure already appears in the original KWT equations. Kramer and Watts-Tobin write their anomalous term as an energy integral of the nonequilibrium distribution function already in their starting equation~\cite{KramerWattsTobin1978}, in a form that plays the role of our full $\mathcal{Q}$ rather than $\mathcal{Q}^{\textrm{hom}}$ or $\mathcal{Q}^{\textrm{part}}$ separately [Eq.~(1) of Ref.~\cite{KramerWattsTobin1978}]. After their local-equilibrium reduction, this term becomes a contribution proportional to $\partial|\psi|^2/\partial t$ within the same time-derivative bracket as the ordinary TDGL relaxation term [Eq.~(15) of Ref.~\cite{KramerWattsTobin1978}, restated as Eq.~(196) of Ref.~\cite{watts1981nonequilibrium}]. This is exactly the structure we obtain above for $\mathcal{Q}^{\textrm{part}}$ in the short-$\lambda_{\textrm{E}}$ limit. Their setting, however, is a current-driven filament at a single, uniform temperature, so neither a boundary temperature difference nor a counterpart to $\mathcal{Q}^{\textrm{hom}}$ enters their problem.

Combining the homogeneous contribution absorbed into $T_{\textrm{eff}}$ with the short-$\lambda_{\textrm{E}}$ form of the particular contribution, the TDGL equation can be summarized as
\begin{align}
&\left(\gamma_1 + \frac{\tau_{\textrm{E}} }{\epsilon} \int_0^\infty d\omega\, \frac{R_2^2}{N_1} \frac{df^{(0)}_{\textrm{L}}(\omega)}{d \omega}\right)\frac{\partial \Delta}{\partial t} \notag \\
&\quad = \xi^2\frac{\partial^2}{\partial x^2} \Delta + \left(  -\frac{T_{\textrm{eff}}(x,t)-{T_{\textrm{c}}}}{T_{\textrm{c}}-T_{\textrm{L}}} - \frac{|\Delta|^2}{\Delta_{\infty}^2} \right) \Delta.
\end{align}

\subsection{Comparison to the previous work}
\subsubsection{Comparison to Ref.~\cite{Kanakubo2026}}
This work extends Ref.~\cite{Kanakubo2026} by replacing its phenomenological local-temperature description with an explicit treatment of nonequilibrium quasiparticle dynamics. We summarize the relation between the two studies below. Table~\ref{table_comparison} lists the correspondence between the coupled field(s), the governing equations, and the force-balance structure of the two models. The thermal effect enters through different fields in the two models. In Ref.~\cite{Kanakubo2026}, it is represented by a local temperature $\tau(x,t)$, whereas in the present work it is encoded in $\delta f_{\textrm{L}}(x,t,\omega)$ and $\Theta(x,t,\omega)$. Despite this difference, the two models agree on the direction of DW motion and on the qualitative structure of the force balance between a viscous contribution and a thermally driven contribution. We do not postulate a local temperature or a Fourier-type diffusion equation for it. Instead, we derive the same qualitative behavior from the coupled dynamics of the order parameter, the spectral functions, and the nonequilibrium distribution function, with no assumption of local thermal equilibrium beyond the two boundaries. This microscopic treatment further separates the anomalous term into a boundary-driven part that drives the DW and a motion-induced part that renormalizes its relaxation. Such a separation is absent in the single local-temperature field used in Ref.~\cite{Kanakubo2026}.

\begin{table*}[t!]
\caption{Comparison between the local-temperature TDGL model of Ref.~\cite{Kanakubo2026} and the microscopic model used in the present work.}
\label{table_comparison}
\centering
\renewcommand{\arraystretch}{1.6}
\setlength{\tabcolsep}{10pt}
\begin{tabular}{lll}
\toprule
 & Ref.~\cite{Kanakubo2026} & Present work \\
\midrule
Coupled field(s) & Local temperature $\tau(x,t)=T(x,t)/T_{\textrm{c}}$ & $\delta f_{\textrm{L}}(x,t,\omega)$ and $\Theta(x,t,\omega)$ \\
TDGL equation & Linear coefficient $\propto \tau(x,t)-1$ & Anomalous term $\propto\displaystyle\int_0^\infty\! d\omega\,R_2\,\delta f_{\textrm{L}}$ \\
Equation for the field & Thermal diffusion equation & Usadel Eq.~\eqref{fullRUsadel} + transport Eq.~\eqref{fullBoltzmann} \\
\multicolumn{3}{l}{DW velocity formula} \\
\multicolumn{3}{c}{
\begin{minipage}[t]{0.46\textwidth}\centering\scriptsize
\[
  v=\frac{\displaystyle\int_{x_{\textrm{L}}}^{x_{\textrm{R}}}\!dx\,\dfrac{d\tau_1(x)}{dx}\bigl(\psi_0^2(x_{\textrm{R}})-\psi_0^2(x)\bigr)}{\displaystyle 2(1-\tau_{\textrm{L}}) \int_{x_{\textrm{L}}}^{x_{\textrm{R}}}\!dx\,\Bigl(\dfrac{d\psi_0(x)}{dx}\Bigr)^{\!2}}
\]
Eq.~(43) of Ref.~\cite{Kanakubo2026}
\end{minipage}
\hfill
\begin{minipage}[t]{0.48\textwidth}\centering\scriptsize
\[
  v=\frac{\displaystyle\int_{-L/2}^{L/2}\!dx\,\dfrac{d\mathcal{T}(x)}{dx}\bigl(|\Delta^{(0)}(L/2)|-|\Delta^{(0)}(x)|\bigr)}{\displaystyle\epsilon\gamma_1\!\int_{-L/2}^{L/2}\!dx\Bigl(\dfrac{d\Delta^{(0)}}{dx}\Bigr)^{\!2}+\int_{-L/2}^{L/2}\!dx\,\dfrac{d|\Delta^{(0)}|}{dx}\displaystyle\int_0^\infty\!d\omega\,R_2^{(0)}\tilde\Phi_{\textrm{p}}}
\]
Eq.~\eqref{vselect_TL_alt}
\end{minipage}
} \\[4pt]
DW direction & Hotter side & Hotter side \\
Force balance & Viscous (cold) + thermal force (hot) = 0 & Viscous (cold) + anomalous force (hot) = 0  \\
\bottomrule
\end{tabular}
\end{table*}

\subsubsection{Comparison to Refs.~\cite{deLange1974, Hu1976}}

Because Refs.~\cite{deLange1974,Hu1976} also analyze an anomalous term in an Eliashberg-type TDGL equation for vortex dynamics, it is useful to compare their setting with ours. Table~\ref{table_comparison_deLangeHu} summarizes the main points. The two settings are complementary rather than competing. The driving mechanism and the microscopic origin of the anomalous term both differ between them.

References~\cite{deLange1974,Hu1976} consider an isolated vortex driven by a transport current at a uniform temperature, with an anomalous term originating from pair-breaking by paramagnetic impurities. They compute the transport entropy $S_d$, an Onsager coefficient, from which a thermal force of the kind proposed by Stephen for a vortex line~\cite{Stephen66} follows only through a further Onsager-reciprocity argument. In the present work, by contrast, the temperature difference is imposed directly through the boundary conditions on a DW, the anomalous term originates from inelastic electron--phonon scattering, and the DW velocity is obtained from the force balance derived in Sec.~\ref{sec:sign}.

Despite these differences, both studies identify part of the anomalous term with a local temperature deviation. Neither Ref.~\cite{deLange1974} nor Ref.~\cite{Hu1976} imposes a temperature difference at the boundary of the sample. Their vortex is instead driven by an external transport current at a spatially uniform bath temperature, and their entire anomalous term is generated by the vortex's own motion, through the source $\partial_t|\Delta|^2$. This is the same source that generates our motion-induced part $\mathcal{Q}^{\textrm{part}}$, which vanishes identically when $\partial_t\Delta=0$ [cf. Eqs.~\eqref{eq:delfLhom} and \eqref{eq:delfLpart}]. Section~\ref{sec:direction} shows that this motion-induced part cannot set the sign of the DW velocity, so we build $T_{\textrm{eff}}$ from the boundary-driven part $\mathcal{Q}^{\textrm{hom}}$ instead, which has no counterpart in their transport-current-driven setting.

Neither reference keeps the anomalous term resolved in the quasiparticle energy. De Lange's $U_1$ satisfies a diffusion equation with no explicit energy dependence, and $U_2$ and Hu's $U$ are reported only after integrating over energy [Eqs.~(3), (4) of Ref.~\cite{deLange1974}, Eq.~(10b) of Ref.~\cite{Hu1976}]. Our formalism instead keeps $\delta f_{\textrm{L}}(x,t,\omega)$ resolved in $\omega$ throughout. This resolution shows that quasiparticles near the local gap edge dominate the anomalous force. $R_2^{(0)}(x,\omega)$ peaks at $\omega\sim|\Delta^{(0)}(x)|$ and vanishes below the gap in the BCS limit, as discussed in Sec.~\ref{sec:sign}.

\begin{table*}
\caption{Comparison between Refs.~\cite{deLange1974,Hu1976} and the present work.}
\label{table_comparison_deLangeHu}
\centering
\renewcommand{\arraystretch}{1.6}
\setlength{\tabcolsep}{16pt}
\begin{tabular}{lll}
\toprule
 & Refs.~\cite{deLange1974,Hu1976} & Present work \\
\midrule
Setup & Vortex, transport current & DW, imposed $\nabla T$ \\
Anomalous-term origin & Paramagnetic-impurity pair-breaking & Inelastic electron--phonon scattering \\
Quantity obtained & Transport entropy $S_d$ &DW velocity $v(\Delta T)$ \\
Local-temperature link & Anomalous term $U$ (motion-generated) & $\mathcal{Q}^{\textrm{hom}}$ (boundary-driven) \\
Thermal force derivation & Indirect (Onsager reciprocity) & Direct (force balance relation) \\
\bottomrule
\end{tabular}
\end{table*}

\subsection{Physical implications and extensions}

A central physical implication of this work is the exact separation of the nonequilibrium distribution into a boundary-driven part and a motion-induced part. This separation is a general property of the energy-mode transport equation of the KWT framework, Eq.~\eqref{fullBoltzmann}. The equation is linear in $\delta f_{\textrm{L}}$, so any solution splits this way regardless of the physical scenario. The physical role played by each part, however, is not the same in every problem. In the present problem, Sec.~\ref{sec:direction} shows that the boundary-driven part alone sets the direction of DW motion, while the motion-induced part only renormalizes the relaxation. The boundary-driven part behaves as a diffusive, temperature-like factor, and the motion-induced part, in the appropriate limit, acts as a positive correction to the ordinary TDGL relaxation time. Associating this motion-induced part with an effective heating or cooling of the quasiparticle population agrees with an interpretation reached independently in the study of phase-slip centers~\cite{Vodolazov2010}, where the sign of the analogous term tracks the growth or decay of the order parameter. By contrast, the present boundary-value problem combines a boundary-driven gradient with the DW's own dynamics, allowing the two contributions to be separated and assigned distinct physical roles.

This identification of the boundary-driven part with an effective temperature connects directly to Ref.~\cite{Kanakubo2026}, and it also shows where that connection breaks down. In Sec.~\ref{sec:efftemp} we used $T_{\textrm{eff}}$ only as a diagnostic, to check what a local-temperature picture would predict once imposed on our microscopic result. $T_{\textrm{eff}}$ diverges at the DW center, where the order parameter vanishes. Sec.~\ref{sec:efftemp} traces this divergence to the way $T_{\textrm{eff}}$ is constructed, not to the underlying dynamics. The DW still moves toward the hotter boundary despite this breakdown of the local-temperature picture at its center. This suggests an energetic preference for where the suppressed order parameter is located. $T_{\textrm{eff}}$ is not the only possible definition of a local effective temperature. How temperature should be defined for a system away from equilibrium is a broader problem, outside the scope of the present work.

This is also why the present work solves the coupled boundary-value problem for the order parameter, the spectral functions, and the nonequilibrium distribution function directly, without introducing a local temperature field anywhere in the model. Such a field is difficult to justify microscopically inside a rapidly varying defect such as a DW, where the order parameter itself changes on the scale of the coherence length. The conventional TDGL equation implicitly assumes that the dynamics of such a defect reduces to the dynamics of the order parameter alone. A DW, like a vortex, generally hosts bound states~\cite{stone1996}. These states are quasiparticle degrees of freedom bound to the defect, not part of the order parameter itself, so this assumption need not hold. The present result does not rely on such a reduction.

A result obtained for a DW does not automatically extend to a vortex. Reference~\cite{Kanakubo2026} treated the DW and the vortex side by side within the conventional TDGL equation, and found that their velocity formulas share an analogous structure and predict the same direction of motion, both under a temperature gradient and under a spin-accumulation gradient. The agreement obtained here therefore suggests, though it does not prove, that the same microscopic treatment applied to a vortex would predict the same direction. A full microscopic treatment of the vortex would need to solve Amp\`ere's law self-consistently with the order parameter, extend the geometry to two transverse dimensions, and keep the charge mode coupled to the energy mode, as noted in Sec.~\ref{sec:intro}. A further extension to a spin-accumulation gradient, the corresponding driving mechanism studied phenomenologically in Ref.~\cite{Kanakubo2026}, could be formulated using the four-distribution-function generalization of the energy and charge modes developed for spin- and charge-imbalanced superconductors~\cite{takane2006boltzmann}.

The DW problem also has significance independent of this comparison. The equations of motion used here, Eqs.~\eqref{fullEqs}, are the KWT equations~\cite{KramerWattsTobin1978}. Kramer and Watts-Tobin originally developed them to describe phase-slip centers in current-carrying superconducting filaments. There, a spatially localized region of suppressed order parameter periodically collapses and reforms, driving a large nonequilibrium quasiparticle population. A stationary, temperature-driven DW of the kind studied here and a dynamic phase-slip center are described by the same equations under different boundary and driving conditions. The connection to DW physics is also relevant to the FFLO state, whose basic structure can be viewed as a periodically modulated array of DWs.

This connection to the FFLO state extends further. A closely related, spatially modulated superconducting state can also be stabilized purely by a nonequilibrium electron distribution, without any spatial temperature or spin-accumulation gradient~\cite{kawamura2024emergence,kawamura2025engineering,kawamura2026emergence}. The microscopic treatment of the nonequilibrium distribution function developed in this work therefore applies with little modification to both problems. A similar Keldysh quasiclassical route from a microscopic model to coupled TDGL and Boltzmann transport equations has also been developed independently for a charge-density-wave conductor~\cite{takane2016time}. This indicates that the present approach extends beyond superconductivity to other systems with a complex order parameter. The underlying boundary-value problem, a position-dependent nonequilibrium distribution function set by reservoirs at the sample edges, is broader still. A nonequilibrium Green's function calculation has recently solved the same type of problem for a normal-state nanowire connected to reservoirs at different electrochemical potentials~\cite{kawamura2026nonequilibrium}, using a route independent of the quasiclassical approach taken here.

\section{Summary} \label{sec:summary}
To summarize, we have studied the dynamics of a domain wall (DW) in a type-II superconductor connected to two heat reservoirs. We used a TDGL framework in which the nonequilibrium quasiparticle distribution function is treated as an independent dynamical variable coupled to the order parameter through spectral functions obtained from the Usadel equation. Because the DW order parameter stays real, only the energy
mode of this distribution function contributes. No local or
nonequilibrium temperature field appears anywhere in the model. The reservoir temperatures enter only through the boundary conditions on the distribution function. Numerical solutions of the full nonlinear
equations and a linear-response calculation of the DW velocity both
show that the DW moves toward the hotter boundary.

A central result of this work is that the nonequilibrium distribution separates exactly into a boundary-driven homogeneous component and a DW-motion-induced particular component. These two components play distinct physical roles. The former provides the driving contribution that fixes the direction of motion. The latter modifies the relaxation of the order parameter and reduces, in the strong-inelastic-scattering limit, to a motion-induced correction to the TDGL relaxation term. The local momentum-balance relation then shows that the resulting DW motion is toward the hotter boundary over the wide range of inelastic-scattering rates studied in this work. This result agrees with our previous phenomenological study based on an explicit local temperature field~\cite{Kanakubo2026} and confirms its prediction within a more microscopic nonequilibrium framework. The motion toward the hotter boundary therefore does not rely on the local-temperature approximation used in Ref.~\cite{Kanakubo2026}.

\begin{acknowledgments}
We thank H. Adachi, M. Ichioka, J.-i. Ohe, Y. Takane, G. Tatara and T. Yamaguchi for their helpful comments on our manuscript. T.K. was financially supported by World-Leading
Innovative Graduate Study Program of Advanced Basic Science Course (WINGS-ABC) of the University of Tokyo and JST SPRING, Grant Number JPMJSP2108. This research was supported by JSPS KAKENHI Grants (No. 25K23363 and No. 26K00658). The computation in this work was performed using the facilities of the Supercomputer Center, the Institute for Solid State Physics, the University of Tokyo.
\end{acknowledgments}

\appendix
\section{Sign of $N_1^{(0)}$, $R_2^{(0)}$, and $K^{(0)}$} \label{app:R2sign}

\subsection{Decomposition and $K^{(0)}$}

Writing $\Theta^{(0)}(x,\omega) = a(x,\omega) + ib(x,\omega)$ with $a,b$ real,
\begin{equation} \label{N1R2_ab}
  N_1^{(0)} = \cos a \cosh b, \qquad R_2^{(0)} = \cos a \sinh b.
\end{equation}
The same two relations give
\begin{align} \label{K0_cos2a}
  K^{(0)} &\equiv (N_1^{(0)})^2-(R_2^{(0)})^2 \notag \\
  &= \cos^2 a\,(\cosh^2 b - \sinh^2 b) = \cos^2 a.
\end{align}
$N_1^{(0)}$ is the local density of states, normalized to its normal-state
value. Its non-negativity for a physically sensible retarded Green's
function is a general property, independent of the profile of
$\Delta^{(0)}(x)$, and we do not reprove it here. Because $\cosh b > 0$
for any real $b$, $N_1^{(0)}>0$ is equivalent to $\cos a > 0$.
Eq.~\eqref{K0_cos2a} then gives $K^{(0)}=\cos^2 a>0$ strictly, with no
assumption beyond $N_1^{(0)}>0$. Eq.~\eqref{N1R2_ab} also gives
\begin{align} \label{R2_sign_reduction}
  \mathrm{sgn}\bigl(R_2^{(0)}\bigr) &= \mathrm{sgn}(\sinh b) = \mathrm{sgn}(b) \notag \\
  &= \mathrm{sgn}\bigl(\mathrm{Im}[\Theta^{(0)}]\bigr).
\end{align}
Showing $R_2^{(0)}(x,\omega)>0$ is therefore equivalent to showing
$b(x,\omega)>0$.

Inserting $\Theta^{(0)}=a+ib$ into Eq.~\eqref{0thRUsadel} and separating
real and imaginary parts, with $\sigma\equiv-\hbar/\tau_{\textrm{E}}$ and
$\eta\equiv2\omega$, gives
\begin{align}
  \hbar D\,a'' + \sigma\sin a\cosh b - \eta\cos a\sinh b \notag \\
  + 2|\Delta^{(0)}(x)|\cos a\cosh b &= 0, \label{aeq} \\
  \hbar D\,b'' + \sigma\cos a\sinh b + \eta\sin a\cosh b \notag \\
  - 2|\Delta^{(0)}(x)|\sin a\sinh b &= 0. \label{beq}
\end{align}
We examine the sign of $b$ in three regimes below.

\subsection{Boundary, for $\omega>0$}

At $x=\pm L/2$, $\Theta^{(0)}=\Theta_{\textrm{bulk}}(\omega;T_{\textrm{L}})$.
Write
\begin{gather}
  s\equiv2i\omega-\hbar/\tau_{\textrm{E}}=\sigma+i\eta, \label{sdef}\\
  w\equiv\sqrt{s^2+4\Delta_\infty^2(T_{\textrm{L}})}=w_{\textrm{r}}+iw_{\textrm{i}}, \label{wdef}
\end{gather}
with $w_{\textrm{r}},w_{\textrm{i}}$ real. Eq.~\eqref{0thBCsforTheta} gives $\cos\Theta_{\textrm{bulk}}=s/w$
and $\sin\Theta_{\textrm{bulk}}=-2\Delta_\infty(T_{\textrm{L}})/w$, with the branch of $w$ fixed by the physical condition $\mathrm{Re}[s/w]>0$.

For $\omega>0$, $\eta>0$. Since $\Delta_\infty^2(T_{\textrm{L}})$ is real,
\begin{equation} \label{Imw2}
  \mathrm{Im}[w^2]=\mathrm{Im}[s^2]=2\sigma\eta<0,
\end{equation}
so $w_{\textrm{r}}$ and $w_{\textrm{i}}$ have opposite signs. Only one
of the two roots $\pm w$ satisfies $\mathrm{Re}[s/w]>0$, equivalently
\begin{equation} \label{branchcond}
  \mathrm{Re}[s\bar w]=\sigma w_{\textrm{r}}+\eta w_{\textrm{i}}>0.
\end{equation}
If $w_{\textrm{r}}>0$ and $w_{\textrm{i}}<0$, then $\sigma
w_{\textrm{r}}<0$ and $\eta w_{\textrm{i}}<0$. This root is excluded.
If $w_{\textrm{r}}<0$ and $w_{\textrm{i}}>0$, then $\sigma
w_{\textrm{r}}>0$ and $\eta w_{\textrm{i}}>0$. This is the physical
root, and it satisfies $w_{\textrm{i}}=\mathrm{Im}(w)>0$, for every
$\omega>0$, every finite $\tau_{\textrm{E}}>0$, and every
$T_{\textrm{L}}<T_{\textrm{c}}$.

Since $R_2^{(0)}=\mathrm{Im}[\sin\Theta_{\textrm{bulk}}]
=2\Delta_\infty(T_{\textrm{L}})\,\mathrm{Im}(w)/|w|^2$, this gives
$R_2^{(0)}(\pm L/2,\omega)>0$ for $\omega>0$, exactly.

\subsection{The entire sample, at $\omega=0$}

At $\omega=0$, $\eta=0$ in Eq.~\eqref{sdef}, so $s=\sigma$ is real.
Eq.~\eqref{wdef} then gives $w=\sqrt{\sigma^2+4\Delta_\infty^2(T_{\textrm{L}})}$,
also real, and the branch condition $\sigma/w>0$ with $\sigma<0$ picks
$w<0$. The boundary value $\Theta_{\textrm{bulk}}(0;T_{\textrm{L}})$ is
therefore real.

At $\omega=0$, Eqs.~\eqref{aeq} and \eqref{beq} become
\begin{align}
  \hbar D\,a'' + \sigma\sin a\cosh b 
  + 2|\Delta^{(0)}|\cos a\cosh b = 0, \label{aeq0}\\
  \hbar D\,b'' + \sigma\cos a\sinh b 
  - 2|\Delta^{(0)}|\sin a\sinh b = 0. \label{beq0}
\end{align}
Consider $b(x)\equiv0$ for every $x$. Since $\cosh0=1$, $\sinh0=0$, and
$b''\equiv0$, Eq.~\eqref{beq0} reduces to $0=0$, satisfied identically
for any $a(x)$. This is exact, not an approximation. Eq.~\eqref{aeq0}
reduces to the closed, real ordinary differential equation
\begin{equation} \label{aeq0real}
  \hbar D\,a'' + \sigma\sin a + 2|\Delta^{(0)}(x)|\cos a = 0,
\end{equation}
with the real boundary condition found above, and with the full spatial
dependence of $|\Delta^{(0)}(x)|$ retained.

We have therefore constructed a solution of Eqs.~\eqref{aeq0} and
\eqref{beq0}, exact at every $x$, given by $b\equiv0$ together with $a(x)$
solving Eq.~\eqref{aeq0real}. Assuming, as we do throughout this work,
that the physical solution of this boundary-value problem is unique, this
real solution is $\Theta^{(0)}(x,0)$. It follows that
\begin{equation}
  R_2^{(0)}(x,0)\equiv0
\end{equation}
exactly, throughout the sample. Evaluating $R_2^{(0)}(x,0)$ directly on
the numerical grid of Sec.~\ref{sec_num} gives values with magnitude at
most $2\times10^{-12}$, consistent with floating-point noise around this
exact value.

\subsection{Interior, for $\omega>0$}

Unlike the equation for $\varphi_1$ used in Sec.~\ref{sec:sign}, the
coupled system Eqs.~\eqref{aeq} and \eqref{beq} does not reduce to a
single scalar equation for $b$ alone, for $\omega>0$. The coefficients of
the equation for $b$ depend on $a$, and the equation for $a$ depends on
$b$ in turn. We have not found an analytic argument that closes this
coupled system for general $x$ and $\omega>0$.

\begin{figure}[t!]
\includegraphics[width=\columnwidth]{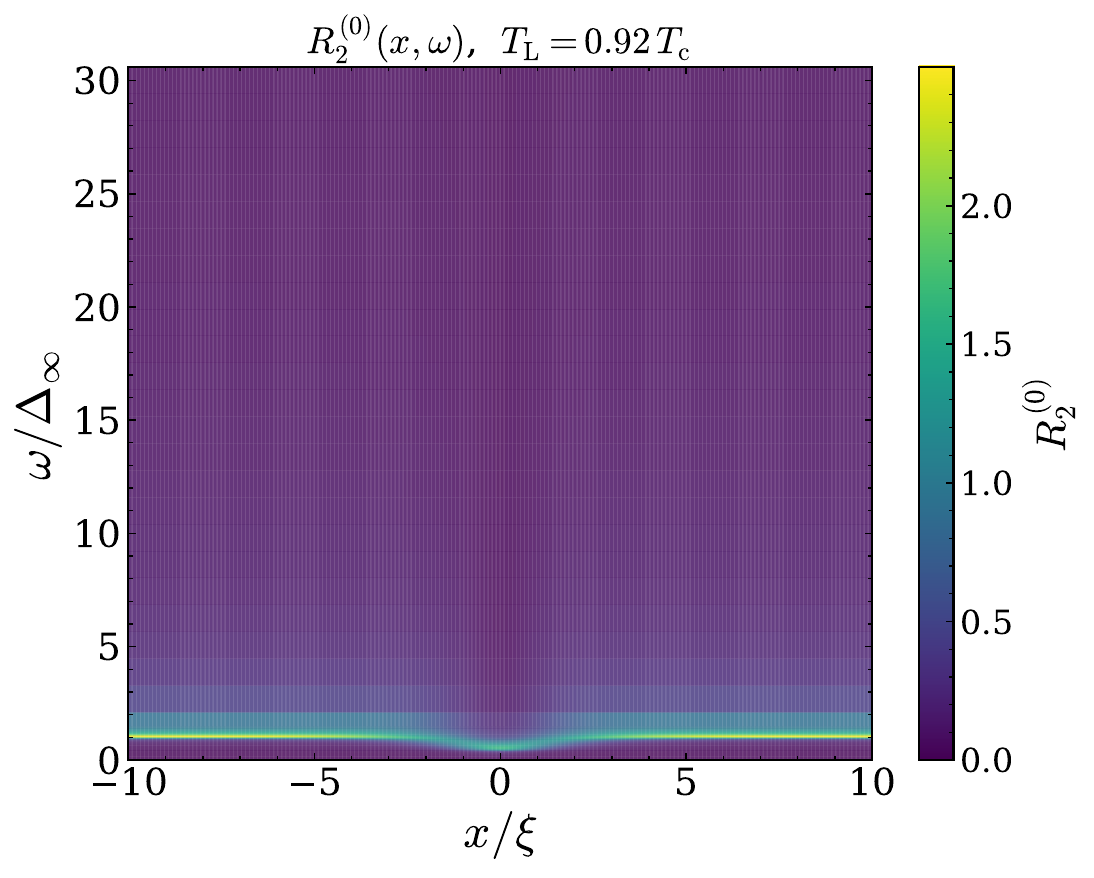}
\caption{Numerical solution of $R_2^{(0)}(x,\omega)$ at
$T_{\textrm{L}}=0.92\,T_{\textrm{c}}$, the parameter set used in
Sec.~\ref{sec_num}. $R_2^{(0)}$ is positive throughout the range shown,
vanishes as $\omega\to0$, and peaks near the coherence peak
$\omega\approx\Delta_\infty$.}
\label{figR2sign}
\end{figure}

We instead evaluated $R_2^{(0)}(x,\omega)$ directly on the numerical
zeroth-order solution used in Sec.~\ref{sec_num} and Sec.~\ref{sec:linear},
at $T_{\textrm{L}}=0.92\,T_{\textrm{c}}$ and $L/\xi=20$. Figure~\ref{figR2sign}
shows the result. $R_2^{(0)}(x,\omega)$ is positive everywhere shown,
peaks near the coherence peak $\omega\approx\Delta_\infty$, and decreases
toward zero as $\omega\to0^+$, consistent with the exact result
$R_2^{(0)}(x,0)=0$ found above. Excluding the boundaries and $\omega=0$,
treated exactly above, the smallest value found on the numerical grid is
\begin{equation}
  R_2^{(0)}(x=9.99\,\xi,\ \omega=0.0211\,\Delta_\infty)=1.053\times10^{-3},
\end{equation}
close to the right boundary and at the smallest sampled frequency. This
is about $8.7$ orders of magnitude above the floating-point noise level found at
$\omega=0$, and it agrees with the boundary value at the same frequency,
$R_2^{(0)}(L/2,\omega)=1.053\times10^{-3}$, to the precision quoted.

We regard the interior positivity of $R_2^{(0)}$, for $\omega>0$, as a
numerically well-supported property of this solution. We do not have a
general proof of it.

\section{Sign of the DW-induced correction to the denominator}
\label{app:GF_sign}

We give here the derivation showing
that the second term in the denominator of the velocity
formula~\eqref{vselect_TL}, namely $I_{\textrm{DW}}$ of
Eq.~\eqref{IDW_main}, is non-negative for all $\tau_{\textrm{E}}>0$.
Section~\ref{sec:sign} summarizes this derivation.

\subsection{Governing equation for $\tilde\Phi_{\textrm{p}}$}

We introduce the source term
\begin{equation}\label{def_Seq_fin}
  S_{\textrm{eq}}(x,\omega)
  \equiv -R_2^{(0)}(x,\omega)\,\frac{df_{\textrm{L}}^{(0)}}{d\omega}\,
  \frac{\partial|\Delta^{(0)}|}{\partial x}.
\end{equation}
$S_{\textrm{eq}}$ is an odd function of $x$. With $K^{(0)}$ defined in
Eq.~\eqref{def_Ltau}, Eq.~\eqref{Phip_def} for $\Phi_{\textrm{p}}$ reads
\begin{equation}\label{Boltz_fin}
  D\frac{\partial}{\partial x}\!\left[K^{(0)}(x,\omega)\frac{\partial\Phi_{\textrm{p}}}{\partial x}\right]
  - \frac{N_1^{(0)}(x,\omega)}{\tau_{\textrm{E}}}\,\Phi_{\textrm{p}}
  = S_{\textrm{eq}}(x,\omega).
\end{equation}
By its definition, Eq.~\eqref{def_tildePhip}, $\tilde\Phi_{\textrm{p}} =
\Phi_{\textrm{p}} - (\Phi_{\textrm{p}}^R/\varphi_1^R)\varphi_1$ solves the
same inhomogeneous equation as $\Phi_{\textrm{p}}$, because $\varphi_1$
solves the homogeneous equation~\eqref{Boltzmann_hom_TL}. Unlike
$\Phi_{\textrm{p}}$, $\tilde\Phi_{\textrm{p}}$ satisfies the
homogeneous Dirichlet condition $\tilde\Phi_{\textrm{p}}(\pm
L/2,\omega)=0$ by construction.

\subsection{Construction of the Green's function}

We define the Green's function $G(x,x';\omega)$ of $ \hat{\mathcal{L}}_\tau$ by
\begin{equation}\label{Greens_fin}
   \hat{\mathcal{L}}_\tau\, G(x,x';\omega) = \delta(x-x'),
  \qquad G(\pm L/2,\,x';\omega) = 0,
\end{equation}
so that
\begin{equation}
\tilde\Phi_{\textrm{p}}(x,\omega) = \int_{-L/2}^{L/2}dx'\,G(x,x';\omega)\,S_{\textrm{eq}}(x',\omega).
\end{equation}

To construct $G$, let $u_-(x,\omega)$ and $u_+(x,\omega)$ denote solutions
of the homogeneous equation $ \hat{\mathcal{L}}_\tau u=0$ satisfying
$u_-(-L/2)=0$ and $u_+(L/2)=0$, respectively. We normalize each to be
positive in the interior $|x|<L/2$. For $x\neq x'$, $G$ solves the
homogeneous equation, so
\begin{equation}\label{GF_form_fin}
  G(x,x';\omega) = \begin{cases}
    A(x';\omega)\,u_+(x,\omega) & (x > x') \\[4pt]
    B(x';\omega)\,u_-(x,\omega) & (x < x')
  \end{cases}
\end{equation}
Continuity at $x=x'$, $A\,u_+(x')=B\,u_-(x')$, together with the
flux-jump condition obtained by integrating Eq.~\eqref{Greens_fin} over
$[x'-\varepsilon,x'+\varepsilon]$ and taking $\varepsilon\to0$,
\begin{equation}\label{jump_fin}
  DK^{(0)}\frac{\partial G}{\partial x}\bigg|_{x=x'^+}
  - DK^{(0)}\frac{\partial G}{\partial x}\bigg|_{x=x'^-} = 1,
\end{equation}
determines $A$ and $B$. The combination $W\equiv
DK^{(0)}(u_-\partial_xu_+-u_+\partial_xu_-)$ is independent of $x$.
Subtracting $u_-$ times the equation for $u_+$ from $u_+$ times the
equation for $u_-$ gives
\begin{equation}\label{abel}
  D\frac{\partial}{\partial x}\!\left[K^{(0)}\!\left(u_-\frac{\partial u_+}{\partial x} - u_+\frac{\partial u_-}{\partial x}\right)\right] = 0,
\end{equation}
so $W$ is constant. Solving the continuity and jump conditions for $A$
and $B$ gives
\begin{equation}\label{coeff_fin}
  A = \frac{u_-(x')}{W}, \qquad B = \frac{u_+(x')}{W},
\end{equation}
and therefore
\begin{equation}\label{GF_explicit_fin}
  G(x,x';\omega) = \frac{u_-(x_<,\omega)\,u_+(x_>,\omega)}{W(\omega)},
\end{equation}
with $x_<\equiv\min(x,x')$ and $x_>\equiv\max(x,x')$.

\subsection{Negative definiteness of $ \hat{\mathcal{L}}_\tau$ and of $G$}

$ \hat{\mathcal{L}}_\tau$ is Hermitian under the Dirichlet condition. For any
$u,v$ satisfying $u(\pm L/2)=v(\pm L/2)=0$, integration by parts gives
\begin{align}\label{selfadj}
  &\int_{-L/2}^{L/2}dx\,u\, \hat{\mathcal{L}}_\tau v \notag \\
  &= -D\int_{-L/2}^{L/2}dx\,K^{(0)}\frac{\partial u}{\partial x}\frac{\partial v}{\partial x} - \frac{1}{\tau_{\textrm{E}}}\int_{-L/2}^{L/2}dx\,N_1^{(0)}\,u\,v \notag \\
  &= \int_{-L/2}^{L/2}dx\,v\, \hat{\mathcal{L}}_\tau u,
\end{align}
and the right-hand side is symmetric under $u\leftrightarrow v$. Setting
$u=v$,
\begin{align}\label{neg_def}
  &\int_{-L/2}^{L/2}dx\,u\, \hat{\mathcal{L}}_\tau u \notag \\
  &= -D\int_{-L/2}^{L/2}dx\,K^{(0)}\!\left(\frac{\partial u}{\partial x}\right)^{\!2}
     - \frac{1}{\tau_{\textrm{E}}}\int_{-L/2}^{L/2}dx\,N_1^{(0)}\,u^2 \notag \\
  &\leq 0,
\end{align}
because $K^{(0)}\geq0$, $N_1^{(0)}\geq0$, and $\tau_{\textrm{E}}>0$, with
equality only for $u\equiv0$. Hence $ \hat{\mathcal{L}}_\tau$ is negative
definite. Being Hermitian, its eigenvalues $\lambda_n$ are real, and
negative definiteness gives $\lambda_n<0$ for all $n$. Expanding the
Green's function in the eigenfunctions $\psi_n$ of $ \hat{\mathcal{L}}_\tau$,
$G(x,x';\omega)=\sum_n\psi_n(x)\psi_n(x')/\lambda_n$, we obtain for any
function $f$
\begin{equation}\label{GFF_neg}
\begin{split}
  &\int_{-L/2}^{L/2}dx\int_{-L/2}^{L/2}dx'\,f(x)\,G(x,x';\omega)\,f(x') \\
  &\quad= \sum_n \frac{1}{\lambda_n}
    \left(\int_{-L/2}^{L/2}dx\,f(x)\,\psi_n(x)\right)^{\!2} \\
  &\quad\leq 0,
\end{split}
\end{equation}
where every term is non-positive because $\lambda_n<0$ and the squared
bracket is non-negative. This establishes that the quadratic form of $G$
is negative semi-definite for any function $f(x)$, that is,
Eq.~\eqref{GFF_neg_main}.

\subsection{Evaluation of $I_{\textrm{DW}}$}

Substituting the Green's-function representation of $\tilde\Phi_{\textrm{p}}$
and the definition of $S_{\textrm{eq}}$, Eq.~\eqref{def_Seq_fin}, into
Eq.~\eqref{IDW_main} gives
\begin{align}\label{IDW_bilin_fin}
  I_{\textrm{DW}}
  &= \int_{-L/2}^{L/2}dx\,\frac{\partial|\Delta^{(0)}|}{\partial x}
     \int_0^\infty d\omega\,R_2^{(0)} \notag \\
  &\quad\times \int_{-L/2}^{L/2}dx'\,G(x,x';\omega)\,S_{\textrm{eq}}(x',\omega) \notag \\[4pt]
  &= -\int_0^\infty d\omega\,\frac{df_{\textrm{L}}^{(0)}}{d\omega}
     \int_{-L/2}^{L/2}dx \notag \\
  &\quad\times \int_{-L/2}^{L/2}dx'\,
     h(x,\omega)\,G(x,x';\omega)\,h(x',\omega).
\end{align}
Since $df_{\textrm{L}}^{(0)}/d\omega = (2k_{\textrm{B}}T)^{-1}\mathrm{sech}^2(\omega/2k_{\textrm{B}}T)
> 0$, and Eq.~\eqref{GFF_neg} gives $\iint dx\,dx'\,h\,G\,h \leq 0$, we
conclude $I_{\textrm{DW}} \geq 0$ for all $\tau_{\textrm{E}}>0$. Sec.~\ref{sec:sign} uses this result.

\bibliography{ref}

\end{document}